\documentclass[aps,pra,noshowpacs,twocolumn,floatfix]{revtex4-1}
\usepackage{graphicx}
\usepackage{epstopdf} 
\usepackage{amsmath}
\usepackage{amsfonts}
\usepackage{mathtools}
\usepackage{braket}
\usepackage{cases}
\usepackage{ulem}
\usepackage{dsfont}
\usepackage{color}
\usepackage{nccmath}
\usepackage{bm}

\begin{document}

\title{Quantum caustics and the hierarchy of light cones in quenched spin chains}

\date{\today}
\author{W. Kirkby}
\affiliation{Department of Physics and Astronomy, McMaster University, 
1280 Main St. W., Hamilton, Ontario, Canada, L8S 4M1}
\author{J. Mumford}
\affiliation{Department of Physics and Astronomy, McMaster University, 
1280 Main St. W., Hamilton, Ontario, Canada, L8S 4M1}
\author{D.H.J. O'Dell}
\affiliation{Department of Physics and Astronomy, McMaster University, 
1280 Main St. W., Hamilton, Ontario, Canada, L8S 4M1}

\begin{abstract}
We show that the light cone-like structures that form in spin chains after a quench are quantum caustics. Their natural description is in terms of catastrophe theory and this implies: 1) a hierarchy of light cone structures corresponding to the different catastrophes; 2)  dressing by characteristic wave functions that obey scaling laws determined by the Arnol'd and Berry indices; 3)  a network of vortex-antivortex pairs in space-time inside the cone. We illustrate the theory by giving explicit calculations for the transverse field Ising model and the XY model, finding fold catastrophes dressed by Airy functions and cusp catastrophes dressed by Pearcey functions; multisite correlation functions are described by higher catastrophes such as the hyperbolic umbilic.  Furthermore, we find that the vortex pairs created inside the cone are sensitive to phase transitions in these spin models with their rate of production being determined by the dynamical critical exponent. More broadly, this work illustrates how catastrophe theory can be applied to singularities in quantum fields.
\end{abstract}

\maketitle

\section{Introduction}

According to Lieb and Robinson \cite{LiebRobinson}, there is a maximum speed $v_{\text{\tiny{LR}}}$ at which information can propagate in discrete quantum systems that obey the Schr\"{o}dinger equation and have short range interactions. This is a powerful and generic statement because it implies that, despite the fact there is no intrinsic speed limit in the (non-relativisitic) Schr\"{o}dinger equation, the response of these many-particle systems to a sudden quench should be in terms of a light cone-like time evolution of spatial correlations \cite{eisert15}. Physically, the ``light cone'' arises from the maximum group velocity of quasiparticles that are excited by the quench and that subsequently propagate through the sample  \cite{Calabrese2006}. Sophisticated methods of analysis have been applied to these quench problems including conformal field theory and tensor networks
\cite{Calabrese2006,Calabrese2005,Lauchli2008,Manmana2009,Stephan2011,Barmettler12,Calabrese2012,Schachenmayer2013,Eisert2013,Hauke2013,Haegeman16,Perfetto17,Kormos17,Cevolani2017}, and the theory has been tested in experiments on ultracold atoms \cite{Cheneau2012,Fukuhara2013,Langen13} and ions \cite{Richerme14,Jurcevic2014} where quantum spin models \cite{Simon2011,Islam2011,Fukuhara2013,Kim2010,Struck2011}, the Bose-Hubbard (BH) model \cite{Jaksch1998,Greiner2002,Stoferle2004,Trotzky2012}, 1D systems \cite{Kinoshita06,Hofferberth07,Jacqmin11}, and quantum walks on a lattice \cite{Karski2009,Preiss2015} can all be realized. The long coherence times of atomic systems make them particularly suited to studying such dynamics \cite{Morsch2006,Blatt12}, and the ability to perform single-site manipulation and detection \cite{Monroe13,Bakr2009,Weitenberg2011,Sherson2010} has enabled unprecedented preparation and visualization of the relevant local observables.

In this paper we show that light cones in quenched spin chains are \textit{quantum caustics}. These are quantum versions of wave focusing phenomena that occur widely in nature in the form of rainbows \cite{Nye}, ship wakes \cite{Kelvin1905,Ursell1994,NIST}, tsunamis and tidal bores \cite{Berry05}, and Cherenkov radiation \cite{Ginzburg05} (including superfluid analogs \cite{Carusotto06,Gladush08,Marino17}). In the geometric ray theory caustics occur where two or more rays coalesce, giving regions in space where the intensity diverges.  By virtue of their singular nature, the natural mathematical description of caustics is via catastrophe theory which partitions them into a hierarchy of equivalence classes, each of which is structurally stable and has its own set of scaling relations \cite{thom75,arnold75,BerryLesHouches}. To show specifically how this approach can be applied to spin chains we consider the exactly solvable 1D XY model \cite{Lieb1961,Katsura1962}, as well as the special case of the 1D transverse-field Ising model (TFIM) \cite{deGennes1963,Pfeuty1969}. While both cases display light cone-like behaviour, the more general XY model allows for an anisotropic coupling giving rise to a double cone \cite{Happola2012,Langen2018}. Although we limit our calculations to these exactly solvable models, the structural stability of catastrophes (insensitivity to small perturbations) guarantees they must survive in the presence of weak non-integrability. This includes weak interactions between quasiparticles or disorder and therefore our results also apply to more general systems than just exactly solvable models.

Wave interference softens caustics and leads to structure on three scales \cite{BerryLesHouches}: at large scales we see divergent ray caustics, whereas at wavelength scales interference smoothes the divergences and dresses each caustic with a characteristic wave function which in the simplest case of two coalescing rays is the Airy function, and finally at subwavelength scales there are networks of vortex-antivortex pairs.  These robust features, including vortex-antivortex networks,  have been observed in optical fields \cite{Nye}, and more recently in electron microscopy \cite{Petersen2013}. They have also been discussed theoretically in the context of Bose-Einstein condensates \cite{Chalker2009,Simula2013} and various aspects seen experimentally in these systems \cite{Rooijakkers2003,Huckans2009,Rosenblum2014}. Furthermore, the association between the Airy function (and its related kernels) and light cones has previously been noted by various authors \cite{Barmettler12,Perfetto17,Kormos17,Najafi2018,Eisler2013,Viti2016,Eisler2018,Hunyadi2004,Allegra2016}, and recent work has conjectured similar universal forms for wavefronts of out-of-time-ordered correlators \cite{Swingle2018,Lin2018,Riddell2019a,Riddell2019b} by examining asymptotic limits of the Airy function. However, to the best of our knowledge the present paper is the first to study the hierarchy of universal wave functions that dress light cones, of which the Airy function is only the first, and also point out that light cones should generically contain networks of vortices which in the case of 1D chains appear as space-time vortices.

\begin{table*}[t]
	\begin{tabular}{|c|c|c|c|c|c|}\hline
		\multicolumn{4}{|c}{\textbf{Generating Function}} & \multicolumn{2}{|c|}{\textbf{Scaling Exponents}} \\\hline
		\textbf{Catastrophe} & $n$ & $Q$ & $\Phi_{Q}(\mathbf{s};\mathbf{C})$ & $\beta_Q$ & $\{\varsigma_m\}$\\\hline
		Fold & 1 & 1 & $s^3/3+Cs$&1/6 & $\varsigma=2/3$\\\hline
		Cusp & 1 & 2 & $s^4/4 + C_2s^2/2 + C_1s$&1/4 &$\varsigma_1=3/4$, $\varsigma_2=1/2$\\\hline
		Swallowtail & 1 & 3 & $s^5/5+C_3s^3/3+C_2s^2/2+C_1s$& 3/10&$\varsigma_1=4/5$, $\varsigma_2=3/5$, $\varsigma_3=2/5$\\\hline
		Butterfly & 1 & 4 & $s^6/6+C_4s^4/4+C_3s^3/3+C_2s^2/2+C_1s$&1/3 &$\varsigma_1=5/6$, $\varsigma_2=2/3$, $\varsigma_3=1/2$, $\varsigma_4=1/3$\\\hline
		Hyperbolic Umbilic & 2 & 3 & $s_1^3/3+s_2^3/3+C_3s_1s_2+C_2s_2+C_1s_1$&1/3 &$\varsigma_1=2/3$, $\varsigma_2=2/3$, $\varsigma_3=1/3$\\\hline
		Elliptic Umbilic & 2 & 3 & $3s_1^2s_2-s_2^3+C_3(s_1^2+s_2^2)+C_2s_2+C_1s_1$&1/3 &$\varsigma_1=2/3$, $\varsigma_2=2/3$, $\varsigma_3=1/3$\\\hline
		Parabolic Umbilic & 2 & 4 & $s_2^4+s_1^2s_2+C_4s_2^2+C_3s_1^2+C_2s_2+C_1s_1$&3/8 &$\varsigma_1=5/8$, $\varsigma_2=3/4$, $\varsigma_3=1/2$, $\varsigma_4=1/4$\\\hline
	\end{tabular}
	\caption{The seven elementary catastrophes and their generating functions $\Phi_{Q}(\mathbf{s};\mathbf{C})$, organized by corank $n$, and dimension $Q$ of control space \cite{Berry1980a}. The associated Arnol'd exponents $\beta_Q$ and Berry exponents $\varsigma_m$ governing the scaling of the wave catstrophes' amplitudes and phase, respectively, are also listed.}
	\label{tab:catastrophetable}
\end{table*}

A fourth scale appears in quantum fields due to discretization of excitations leading to `quantum catastrophes' \cite{leonhardt02,berry04,berry08,odell12,Mumford2017,Mumford2019} (rippling mirrors give analogous effects \cite{Berry75}). Going to the continuum (classical field) limit returns us to a wave catastrophe.  As we shall show, light cones in spin chains have all the features of quantum catastrophes, including \textit{discretized} versions of wave catastrophes and vortices which are regulated by the lattice constant. Although the cone itself is mildly affected by the presence of a quantum critical point (QCP) in the spin models we study,  we find by contrast that the vortices are strongly affected and we use this feature to extract the dynamical critical scaling. 


The rest of this paper is organized as follows: In Sec.\ \ref{sec:Catastrophes} we outline the relevant aspects of catastrophe theory, emphasizing the hierarchy of structures and their scaling properties. In Sec.\ \ref{sec:LightCone} we show that light cones are in fact (quantum) caustics and hence their natural mathematical description is via catastrophe theory. In Sec.\ \ref{sec:XYTFIM} we introduce the XY and TFIM spin chains focusing on the quasiparticle dispersion relation which is the key ingredient we need to apply catastrophe theory. This program is implemented in Sec.\ \ref{sec:pearcey_airy_fns} where we obtain the Airy and Pearcey functions for the wavefunctions dressing the fold and cusp catastrophes/cones in these models. In Sec.\ \ref{sec:scaling} we verify the self-similar scaling properties of light cones that catastrophe theory predicts and in Sec.\ \ref{sec:Correlation} we describe how higher order catastrophes arise in the context of correlation functions. In Sec.\ \ref{sec:Vortices} we identify and discuss the presence of vortex-antivortex pairs within light cones, while in Sec.\ \ref{sec:expt} we touch on the relevance of the theory to quench experiments, and in Sec.\ \ref{sec:conclusion} we conclude with a discussion of the broader significance of the results. In order to make this paper self-contained we have included in appendices \ref{Appdx:BogoliubovDynamics}--\ref{Appdx:vortexscaling} the specifics of quantum spin chain diagonalization methods and various other details of our calculations.

\section{\label{sec:Catastrophes}Geometric and Wave Catastrophes}

In what follows we will not need the full mathematical machinery behind catastrophe theory, but we will make use of a number of key results and for this reason we give a brief overview here. Our treatment is informal, but we emphasize that these results can be proved rigorously.  The main idea can be stated simply: catastrophe theory classifies \textit{structurally stable} singularities of functions and shows that such singularities can only take on certain characteristic shapes \cite{thom75}. In up to four dimensions these are Ren\'{e} Thom's seven elementary \textit{catastrophes} which are listed in Table \ref{tab:catastrophetable}.

Each catastrophe arises from two or more coalescing/bifurcating stationary points of its generating function $\Phi_Q$,  the normal forms for which are given in the table. In the physical applications given in this paper $\Phi_Q$ is the action functional and stationary points therefore correspond to classical paths or rays. From an optical/classical mechanics point of view a catastrophe is a caustic, i.e.\ the locus of points where the ray density diverges. 

Thom's theorem states that the local behaviour of a function near coalescing stationary points can always be mapped by a smooth change of variables onto one of the catastrophes and in this sense catastrophes are universal.  There is also a second sense in which catastrophes are universal:  structural stability means stability against perturbations and thus catastrophes do not require special symmetry and hence occur generically in nature. Perturbations do not qualitatively change catastrophes and only quantitatively affect behaviour up to the strength of the perturbation.

The catastrophes in Table \ref{tab:catastrophetable} are organized by the number $n$ of state variables (their corank), and by the dimension $Q$ of the control parameter space. Control space is the space where the function with its singularities actually lives. The control parameters $\mathbf{C}=\{C_{1},C_{2},\ldots \}$ could be space and time coordinates as well as any other parameters. The state variables $\mathbf{s}=\{s_{1},s_{2},\dots \}$ characterize the rays. The simplest catastrophes (the cuspoids) have $n=1$ and their generating functions are polynomials of the form 
\begin{equation}
\Phi_{Q}(s;\mathbf{C})=\frac{s^{Q+2}}{Q+2}+\sum_{m=1}^{Q}\frac{C_{m}s^{m}}{m},
\end{equation}
with up to $Q$ coalescing stationary points. The stationarity condition reads 
\begin{equation}
\frac{\partial \Phi_{Q}}{\partial s}=0
\label{eq:fermat}
\end{equation}
and corresponds physically to Hamilton's principle of stationary action, while caustics arise from coalescing stationary points where the generating function is stationary to higher order  \cite{BerryLesHouches}
\begin{equation}
\frac{\partial^2 \Phi_{Q}}{\partial s^2}=0 \ .
\label{eq:causticcondition}
\end{equation}
 In the examples we provide in subsequent sections, we focus primarily on the fold and cusp catastrophes, as well as a discussion of the hyperbolic umbilic in the context of correlation functions. Folds and cusps are the only structurally stable singularities in the 2D $(x,t)$ control plane where light cones in 1D chains live, while the higher catastrophes (although they may still exist in greater dimensions) can only be projected onto the plane by way of cusps and folds. This property is generic: catastrophes of higher order contain the lower ones \cite{arnold75}. The cusp is the meeting of two fold lines, the swallowtail contains two cusps, and so on. 

The wavefunctions, or wave catastrophes, associated with catastrophes can be obtained in a way analogous to Feynman path integrals by exponentiating the generating function and integrating over all paths,
\begin{equation}
	\Psi_{Q}(\mathbf{C})  \propto \lambda^{n/2} \int_{-\infty}^{\infty} \cdots \int_{-\infty}^{\infty}  \mathrm{d}^{n}s\;\mathrm{e}^{\mathrm{i} \lambda \, \Phi_{Q}(\mathbf{s};\mathbf{C}) } \ ,
	\label{eq:generalwavecatastrophe}
 \end{equation}
 where $\lambda$ plays the role of the wavenumber $k$ or $1/\hbar$ in quantum problems. In this form, the fact that the generating function plays the role of the physical action becomes clear. These functions are also known as diffraction integrals and many of their properties have been tabulated \cite{NIST}. We emphasize that standard approximations such as the method of stationary phase where the integral over $\mathbf{s}$ is broken up into a sum of independent gaussian integrals around each of the stationary points are doomed to failure when the stationary points coalesce. One must instead keep the full form of $\Phi_{Q}$ to get a result which is uniformly correct through the coalescence regions and this is precisely why diffraction integrals are crucial for treating bifurcation problems where solutions appear or disappear.

\begin{figure}[t]
	\includegraphics[width=0.7\columnwidth]{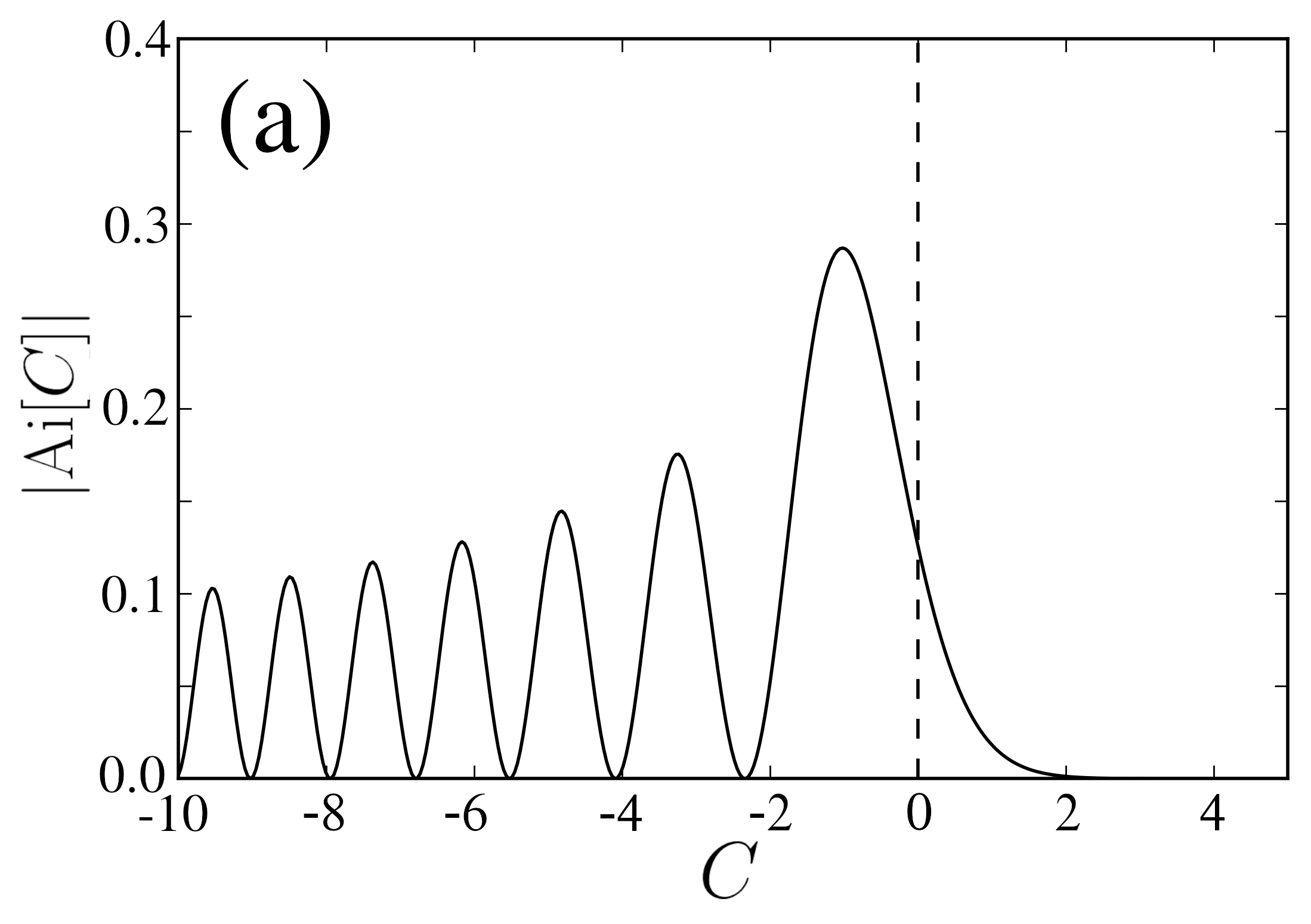}
	\includegraphics[width=0.76\columnwidth]{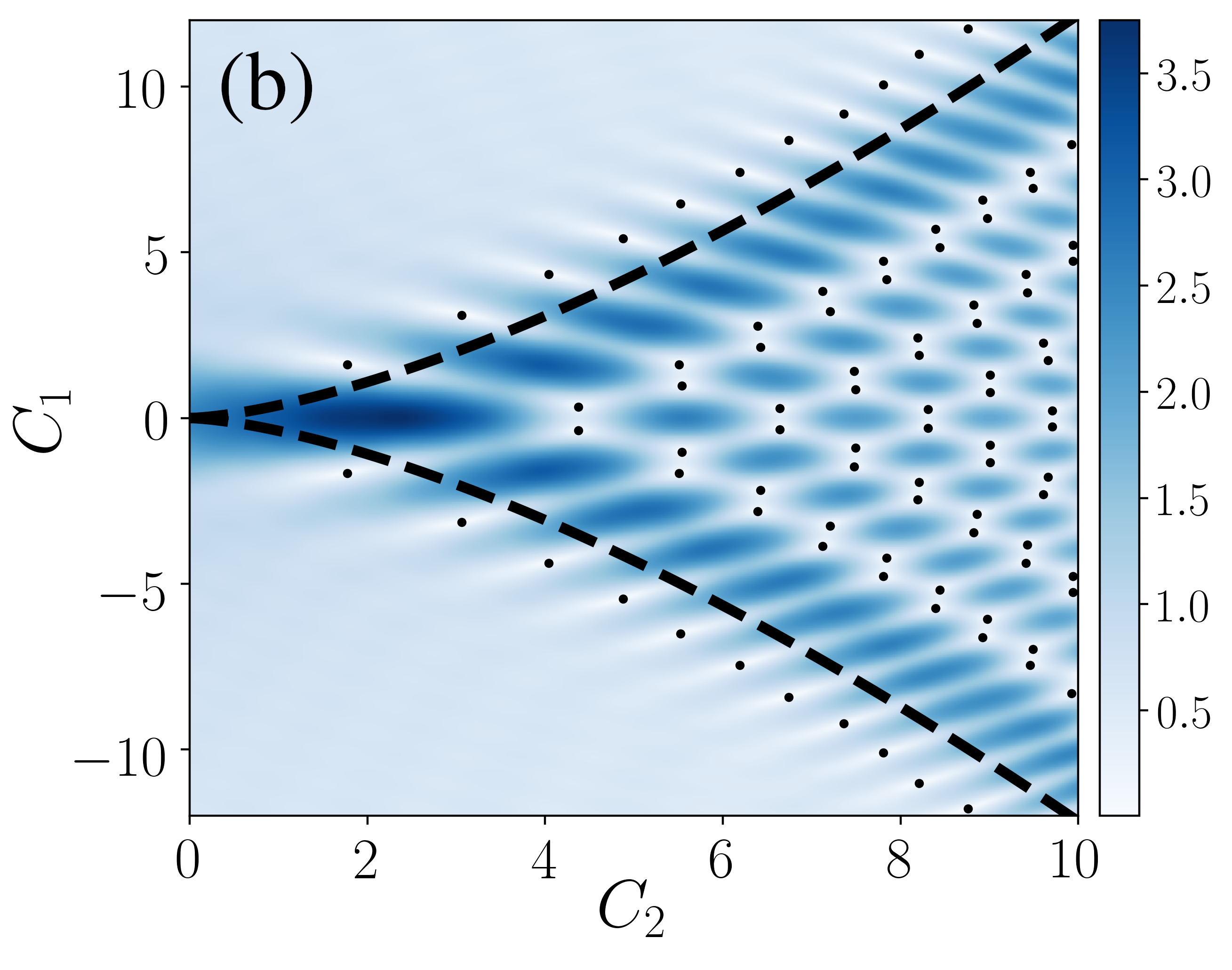}
	\caption{The Airy and Pearcey functions are the first two wave catastrophes in a hierarchy. \textbf{Panel (a):} Modulus of the Airy function, as defined in Eq.\ \eqref{eq:Airy}, which dresses a fold catastrophe where two rays coalesce. The location of the fold, or classical caustic, is at $C=0$ and is indicated by the dashed line. For $C<0$ there is two-wave interference giving fringes whereas for $C>0$ there is an evanescent wave. \textbf{Panel (b):} Modulus of the Pearcey function, as defined in Eq.\ \eqref{eq:Pearcey}, which  dresses the classical cusp caustic $C_1=2C_2^{3/2}/(3\sqrt{3})$ and which is shown as a black dashed line.  The cusp is made of two fold lines which meet at the cusp tip at $C_{1}=C_{2}=0$. There are three rays/waves inside the cusp and only one outside: two coalesce as we cross either of the fold lines, but all three coalesce at the cusp tip which is the most singular part of the classical caustic (a ray picture of the cusp can be seen in Fig.\ 2b in \cite{Mumford2019}). However, wave interference removes the classical singularities. The black dots show the locations of vortices: there is a line of vortices outside either edge of the cusp, and vortex-antivortex pairs inside.}
	\label{fig:PearceyAiry}
\end{figure}

The fold has a cubic action $\Phi_{1}(s;C)=s^3/3+Cs$, where in the case of a light cone in (1+1)-dimensions $C=C(x,t)$. As the control parameter $C$ is taken from positive values down through zero the cubic changes its form so as to describe two coalescing rays. The resulting wave catastrophe can be recognized as the integral form of the Airy function,
\begin{equation}
	\Psi_{1}(C) \propto (2 \pi \lambda^{1/6})  \mathrm{Ai}(\lambda^{2/3} C) \ .
\end{equation}
In the absence of any special symmetry, two fold lines generically meet at cusps. In the region near the cusp point the appropriate action is quartic and features two control parameters $\Phi_{2}(s;C_{1},C_{2})=s^4/4+C_{2} s^2/2 +C_{1} s$. This normal form, which formally resembles the Landau free energy for a continuous (2nd order) phase transition, describes the coalescence of up to three rays and results in a wave catastrophe known as the Pearcey function,
\begin{equation}
	\Psi_{2}(C_{1},C_{2}) \propto (2 \pi \lambda^{1/4})  \mathrm{Pe}(C_{1}\lambda^{3/4},C_{2}\lambda^{1/2})
\end{equation}
which is a complex function of two variables. For our definitions/conventions for the Airy and Pearcey functions, see Eqns. \eqref{eq:Airy} and \eqref{eq:Pearcey}, respectively. Plots of the absolute values $\vert \mathrm{Ai}(C) \vert$ and $\vert \mathrm{Pe}(C_{1},C_{2}) \vert$ of the Airy and Pearcey functions are given in Fig. \ref{fig:PearceyAiry}.

The fact that the Pearcey function is a two-dimensional complex function, with an amplitude and a phase at each point, allows for the possibility of vortices. This turns out to be the case: the black dots in Fig.\ \ref{fig:PearceyAiry}(b) show the locations of vortices, or more precisely their cores. There is an ordered network of vortex-antivortex pairs inside the cusp and single rows of vortices lining the outer edges. These are subwavelength features that represent the finest layer of structure of a wave catastrophe. We find the vortices by densely covering the plane with loops around which we integrate the phase of the Pearcey function: loops that contain vortices give a $\pm 2 \pi$ phase change (the vortex cores also correspond to nodes of the Pearcey function, although in principle not all nodes need be vortices).

An important feature of wave catastrophes is that they exhibit self-similar scaling. If the parameter $\lambda$ is changed from $\lambda'$ to $\lambda$ the wavefunctions will retain their functional forms but with rescaled coordinates,
\begin{equation}
	\Psi_{Q}(\left\{C_m\right\};\lambda)=\left(\frac{\lambda}{\lambda'}\right)^{\beta_Q}\Psi_{Q}\left(\left\{\left(\frac{\lambda}{\lambda'}\right)^{\varsigma_m}C_m\right\};\lambda'\right)\;.
\end{equation}
We can understand this scaling as follows: the overall amplitude scales as $\lambda^{\beta_Q}$, where $\beta_Q$ is known as the \textit{Arnol'd index}. The distance between interference fringes is also rescaled, but generally the scale factor is different in each  direction according to $\lambda^{\varsigma_m}$, where $\varsigma_m$ is the \textit{Berry index} associated with coordinate $C_m$.  For the fold wave catastrophe $\beta_{\mathrm{Ai}}=\frac{1}{6}$ and $\varsigma=\frac{2}{3}$,  and for the cusp wave catastrophe $\beta_{\mathrm{Pe}}=\frac{1}{4}$ and $\varsigma=\{\frac{3}{4},\frac{1}{2}\}$. A complete list of Arnol'd and Berry indices for the seven elementary catastrophes is displayed in Table \ref{tab:catastrophetable}. 

The sets of Arnol'd and Berry indices accompanying the different catastrophes are reminiscent of the sets of critical exponents which define universality classes of equilibrium phase transitions. The underlying common cause of this similarity is the presence of singularities and singularities lead to universality. However, we emphasize that in the light cone case we study here, this universality occurs in non-equilibrium dynamics, and thus we have an example of universality in quantum dynamics.

\section{\label{sec:LightCone}Light Cones as Quantum Caustics}

Our approach to the light cone problem is based upon the idea that the build-up of correlations occurs through quasiparticle propagation \cite{Calabrese2006}; this is known to be the case in a broad range of models including the BH, TFIM, and XY models.
The Lieb-Robinson bound can then be expressed in terms of the maximal group velocity of quasiparticles  \cite{Stephan2011,Calabrese2012}
\begin{equation} 
v_{\text{\tiny{LR}}}=\max_k \left\vert \frac{\mathrm{d} \epsilon_k}{\mathrm{d} k} \right\vert
\end{equation}
where $\epsilon_k$ is the dispersion relation for quasiparticles as a function of quasimomentum $k$. It can be seen immediately that 
this result is exactly equivalent to Eqns.\ (\ref{eq:fermat}) and (\ref{eq:causticcondition}) which give the conditions for a caustic (note that here we are implicitly considering real solutions to the caustic conditions; imaginary solutions correspond to \textit{phase velocity} across the cone and are discussed in Appendix \ref{Appdx:TFIMCaustics}. This aspect has also been discussed by Cevolani \textit{et al.} in Ref.\ \cite{Cevolani2017}). From this simple observation it follows that light cones are caustics and hence the results and insights of catastrophe theory can be applied to them.

Let us focus on the case of a local quench where a single quasiparticle is created at position $x=0$ in the middle of a spin chain (we briefly consider weakly nonlocal superpositions of multiple quasiparticles in Sec.\ \ref{sec:expt}, and also in Appendix \ref{Appdx:SpinFlip}).  Time evolving the state with the Hamiltonian $H$, the state vector at time $t$ is
\begin{equation}\label{eq:PsiVector1}
	\ket{\Psi(t)}  =  \mathrm{e}^{-\mathrm{i}Ht/\hbar}b_{x=0}^\dagger\ket{0}_b 
\end{equation}
where $\ket{0}_b$ is the Bogoliubov quasiparticle groundstate and the operator $b_{x}^{\dagger}$ creates a quasiparticle at the site located at position $x$. For the remainder of the paper, we use the subscript `$b$' to distinguish Fock states in the Bogoliubov basis from the Jordan-Wigner basis. Introducing the eigenstates $\ket{k}$ of $H$ we can write this as (see Appendix \ref{Appdx:BogoliubovDynamics} for details)
\begin{equation}
	\ket{\Psi(t)}  =  \frac{\mathrm{e}^{\mathrm{i}\theta(t)}}{\sqrt{N}}\sum\limits_k\mathrm{e}^{-\mathrm{i}\epsilon_k t/\hbar}\ket{k}_b\;, \label{eq:PsiVector2}
\end{equation}
where $N$ is the number of sites, and the phase $\theta(t)\equiv t/(2\hbar)\sum_k\epsilon_k$ is not observable but is included here for completeness. Projecting onto the position basis,  the wavefunction $\Psi(x_{n},t)\equiv \braket{x_{n}|\Psi(t)}$ on the $n^{\mathrm{th}}$ lattice site is 
\begin{equation}\label{eq:PsiLatticeSum}
	\Psi(x_n,t) = \frac{\mathrm{e}^{\mathrm{i}\theta(t)}}{N}\sum\limits_{k_m=-\pi/a}^{\pi/a-\Delta k}\mathrm{e}^{\mathrm{i}\Phi(k_m;x_n,t)}  \ ,
	\end{equation}
where
	\begin{equation}
\Phi(k;x,t)=kx-\epsilon_kt/\hbar \ .
\label{eq:phasefunc}
\end{equation}
In these expressions $n$ is an integer lying in the range $\{-(N-1)/2,...,(N-1)/2\}$, and the separation between momenta in the sum is $\Delta k=2\pi/(aN)$.

\begin{figure*}[t]
	\centering
	\includegraphics[width=0.66\columnwidth]{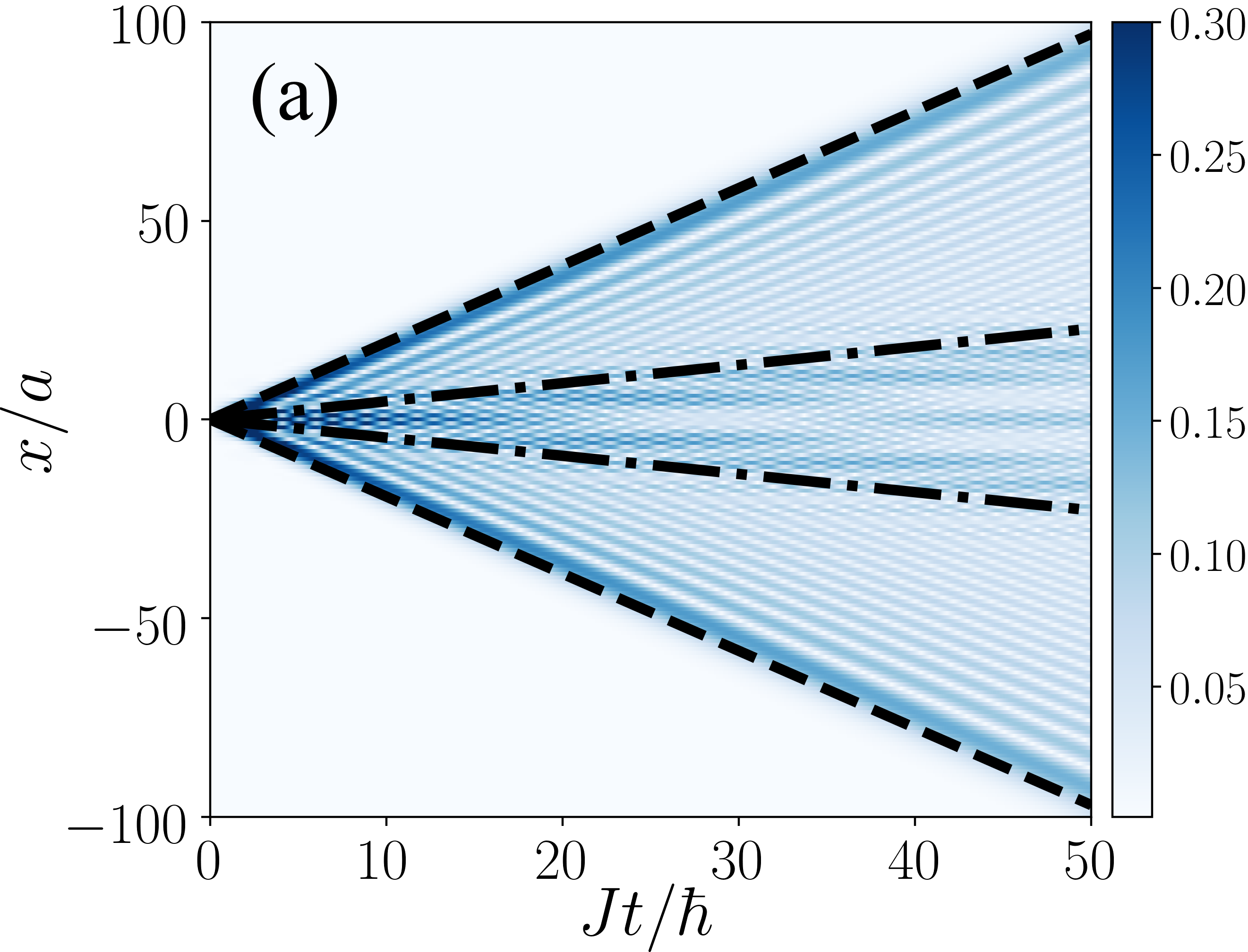}
	\includegraphics[width=0.64\columnwidth]{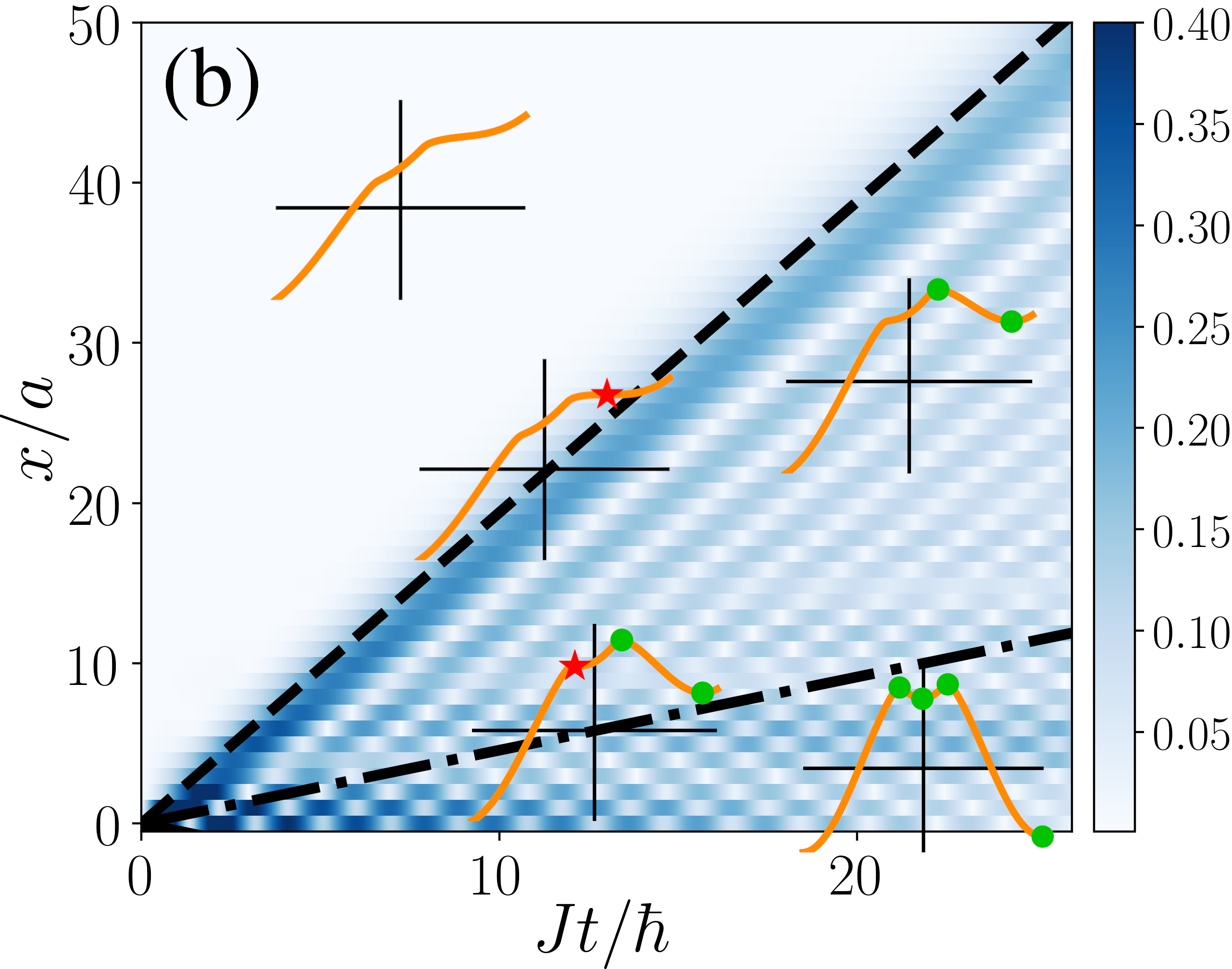}
	\includegraphics[width=0.66\columnwidth]{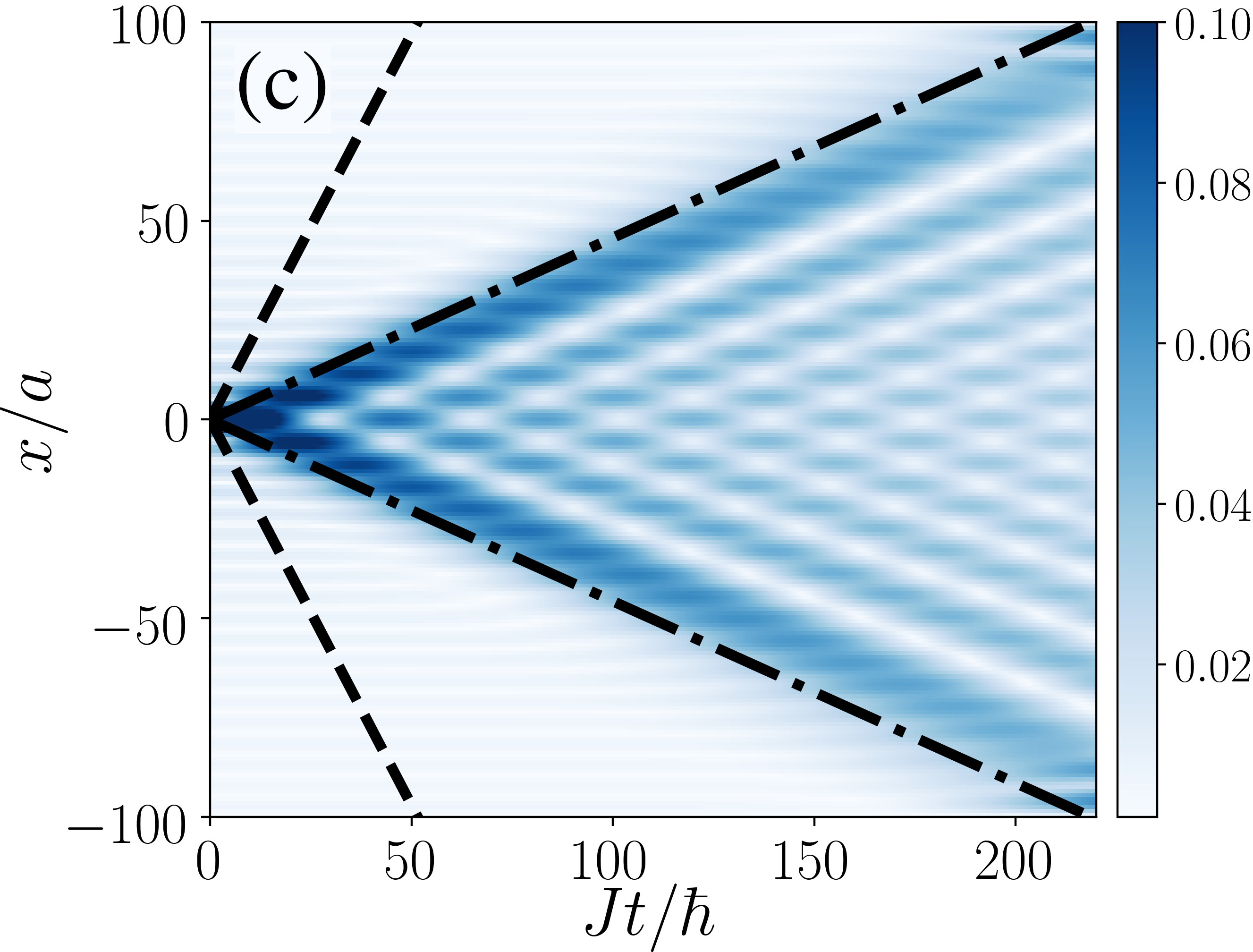}
	\caption{\textbf{Panel (a):}  The exact quantum amplitude, obtained by numerically evaluating Eq.\ \eqref{eq:PsiLatticeSum}, for a single Bogoliubov fermion created at the central lattice site, $x=0$, and propagated under the XY Hamiltonian with $\gamma=0.2$ and $g=0.8$. This model gives rise to a double light cone whose boundaries are indicated by the black dashed (LR cone) and dot-dashed (inner cone) lines. \textbf{Panel (b):} A zoom-in of panel (a) with only half the lattice shown.  At five select points $(x_{n},t)$ we have overlaid plots of the generating function $\Phi(k;x_{n},t)$ [Eq.\ \eqref{eq:Action}] as a function of $k$. Green dots show stationary points of $\Phi(k;x_{n},t)$; there are four stationary points in the inner cone and two annihilate (red stars) each time we cross a cone boundary.  \textbf{Panel (c):} We can isolate the part of $\Psi$ responsible for the inner cone by only including values of $k_{m}$ in Eq.\ \eqref{eq:PsiLatticeSum} that include the three stationary points of $\Phi$ that are close to the centre of the Brillouin zone (note also the change in time scale). As  shown in Sec. \ref{sec:pearcey_airy_fns}, the inner cone is described by a Pearcey function transformed so as to give a straight cone boundary. Note that in order to keep these figures simple we have not shown the vortices although they are present. See Fig.\ \ref{fig:PearceyScaling} below, and also Fig.\ \ref{fig:LightCones2} in the Appendices, for plots of light cone wavefunctions with vortices included.}
	\label{fig:DoubleCone}
\end{figure*}

In the \textit{continuum approximation} (CA) the wavefunction corresponding to Eq.\ \eqref{eq:PsiLatticeSum} is (see Appendix \ref{Appdx:BogoliubovDynamics})
\begin{equation}
	\Psi_{\mathrm{CA}}(x,t)=\frac{\sqrt{a}\;\mathrm{e}^{\mathrm{i}\theta(t)}}{2\pi}\int_{-\pi/a}^{\pi/a} \mathrm{d}k\;\mathrm{e}^{\mathrm{i}\Phi(k;x,t)}\;, 
\label{eq:PsiLatticeInt}	
\end{equation}
where $a=L/N$ is the lattice constant for a lattice of length $L$, and the quasimomentum $k$  runs over the first Brillouin zone. A comparison of the exact (discrete) and CA wavefunctions is given in Fig.\ \ref{fig:LightCones2} in the appendices. In the semiclassical regime, where $N$ is large, the dominant contributions to the integral in Eq.\ (\ref{eq:PsiLatticeInt}) come from values of $k$ where $\Phi$ is slowly varying which are the stationary and coalescence points (especially the latter). By Thom's theorem \cite{thom75,arnold75,BerryLesHouches}, we can therefore map $\Phi$ onto one of the normal forms $\Phi_{Q}$. However, although Thom's theorem guarantees that this can be done by smooth transformations, it does not tell us what these transformations actually are. Figuring out the mapping is part of the challenge in applying catastrophe theory to specific physical problems and it is to this task that we now turn.

\section{XY and TFIM spin chains}
\label{sec:XYTFIM}

Let us consider a 1D XY model describing spins on a lattice interacting with a ferromagnetic coupling $J$, anisotropy parameter $\gamma$, and subject to an external field $gJ$. The Hamiltonian is 
\begin{equation}\label{eq:TFIMHam}
		H=-J\sum_{i}\left(\frac{(1+\gamma)}{2}\sigma_{i}^x\sigma_{i+1}^x+\frac{\left(1-\gamma\right)}{2}\sigma_{i}^y\sigma_{i+1}^y-g\sigma_i^z\right)\;,
\end{equation}
where $\sigma_i^{\alpha}$, $\alpha\in\{x,y,z\}$, are Pauli operators.  When $\gamma=1$ this Hamiltonian reduces to that of the TFIM. The XY Hamiltonian can be diagonalized via the Jordan-Wigner transform followed by a Bogoliubov rotation, which maps spin operators to spinless fermions \cite{Sachdev2011}. As shown in Appendix \ref{appdx:diagonalization}, this leads to the free model $H=\sum_k \epsilon_k (\tilde{b}_k^\dagger\tilde{b}_k-1/2)$, where $\tilde{b}_k^{(\dagger)}$ is the annihilation (creation) operator for Bogoliubov modes with quasimomentum $k$ and dispersion 
\begin{equation}
\epsilon_k=2J\sqrt{(\cos(ka)-g)^2+\gamma^2\sin^2(ka)}  \ . 
\end{equation}
Thus, the phase/generating function in Eq.\ (\ref{eq:phasefunc}) takes the specific form
\begin{equation}
\Phi(k;x,t) =  kx-\frac{2Jt}{\hbar} \sqrt{(\cos(ka)-g)^2+\gamma^2\sin^2(ka)} . \label{eq:Action}
\end{equation}


An exact numerical evaluation of the wavefunction given in Eq.\ (\ref{eq:PsiLatticeSum})  using the generating function $\Phi(k;x,t)$ for the XY model is plotted in Fig.\ \ref{fig:DoubleCone}. The fact that $x_{n}$ is discrete means that the light cone actually corresponds to a  quantum catastrophe,  for more discussion of quantum catastrophes in a spin context see Ref. \cite{Mumford2019}. However, in the semiclassical regime where $N$ is large, the CA described by Eq.\ \eqref{eq:PsiLatticeInt} works well. In this case $\Phi$ has the same functional form but with $x$ and $k$ taken as continuous variables, and the integral can be evaluated analytically in terms of Airy and Pearcey functions as will be explained in the next section.

\begin{figure*}[t]
	\centering
	\includegraphics[width=0.66\columnwidth]{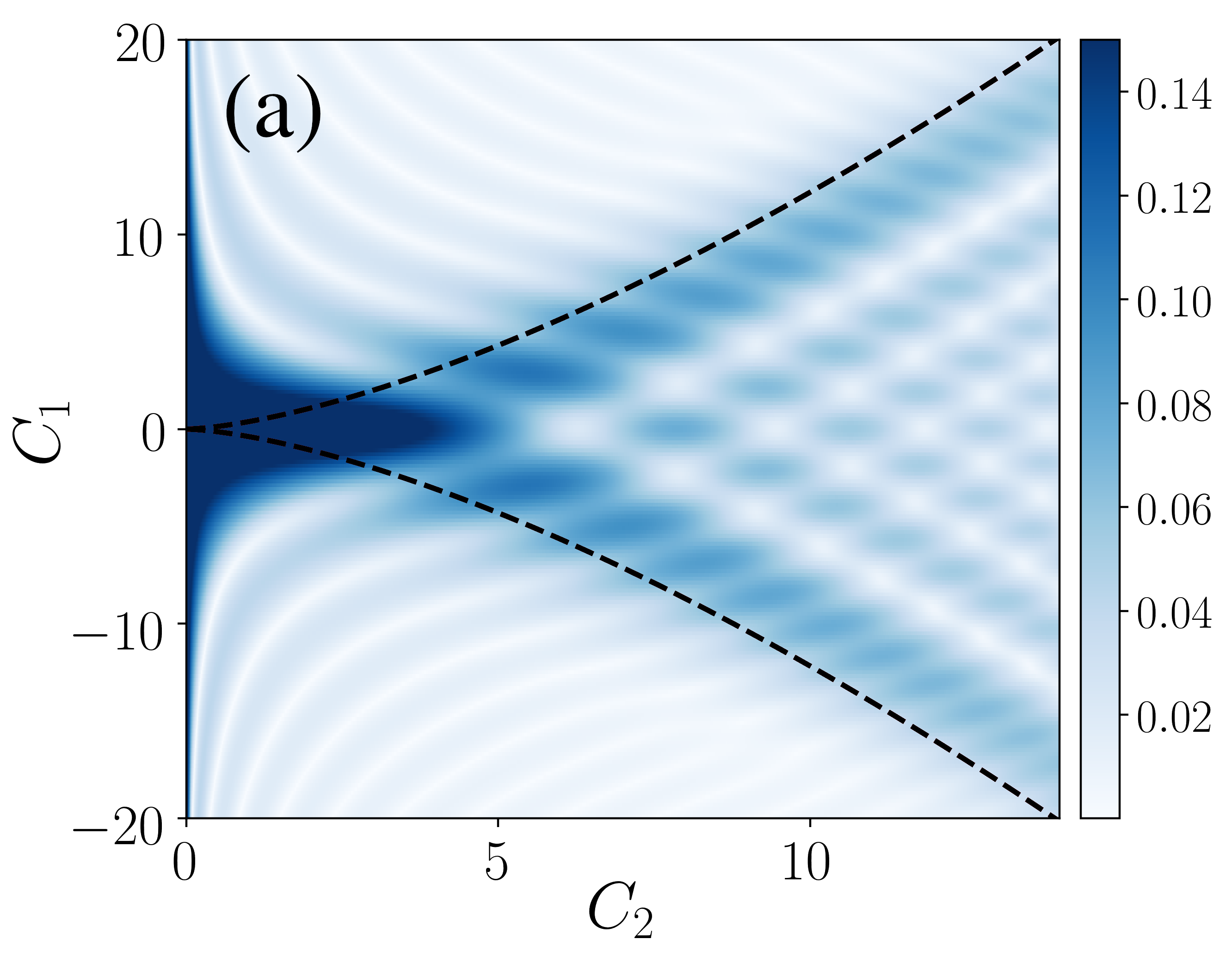}
	\includegraphics[width=0.66\columnwidth]{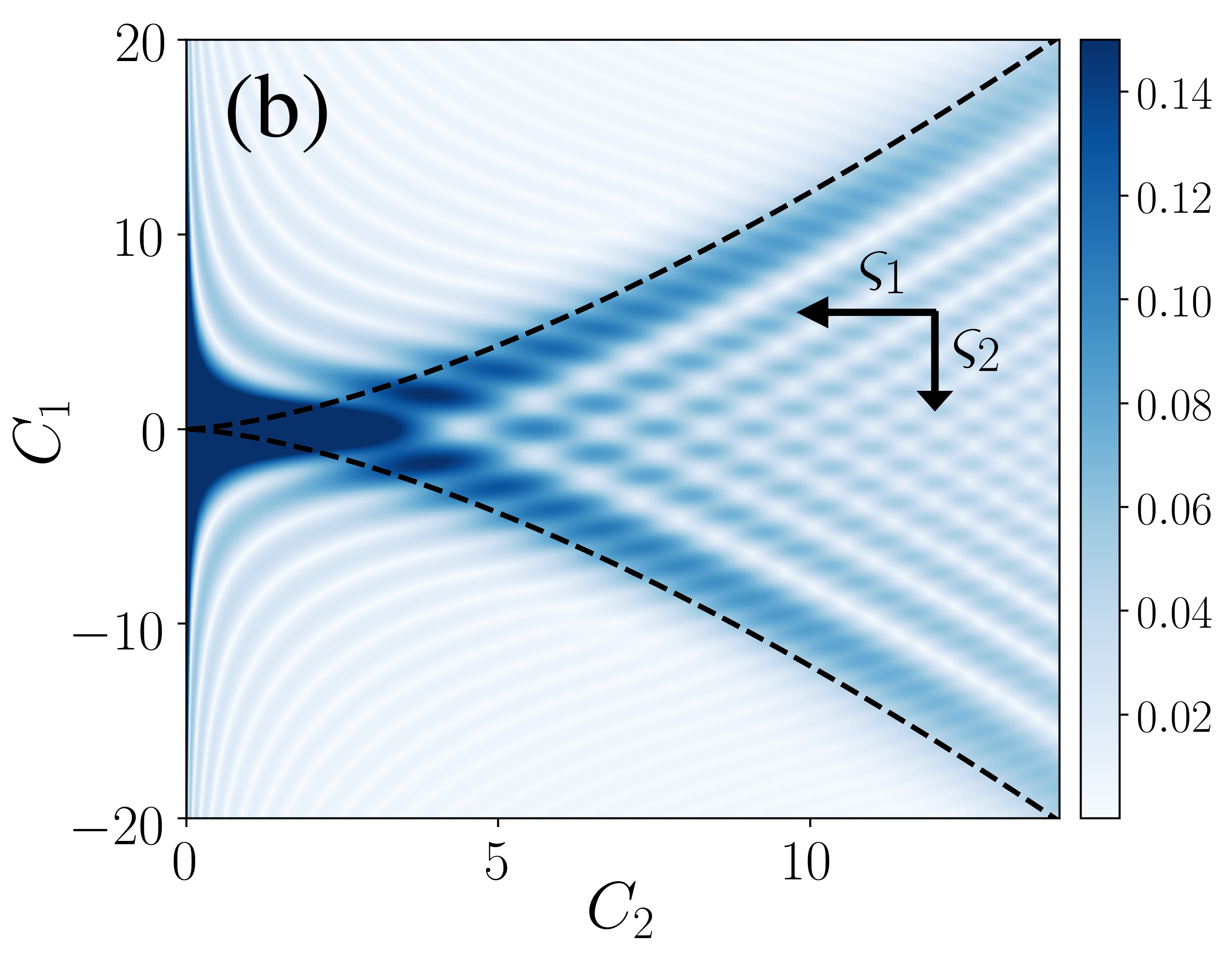}
	\includegraphics[width=0.65\columnwidth]{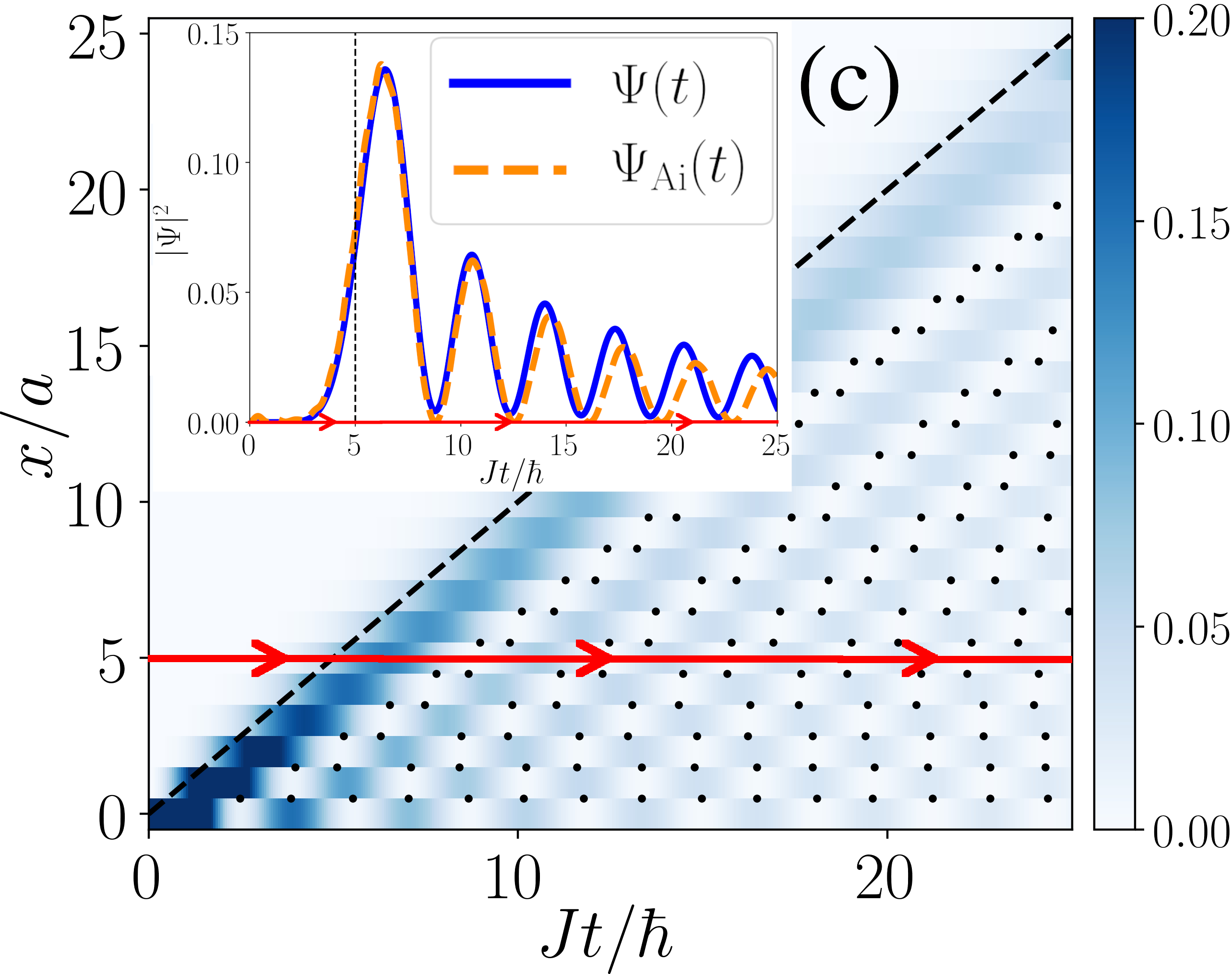}
	\caption{\textbf{Panels (a) and (b):} Modulus of exact wavefunction \eqref{eq:PsiLatticeSum} $\Psi[x(C_1,C_2),t(C_1,C_2)]$ plotted in the scaled coordinates for the inner cone only (vortices are present but not shown). It shows a remarkable qualitative resemblance to the Pearcey function (compare with Fig.\ \ref{fig:PearceyAiry}), without performing any approximations. Between panels (a) and (b), the interaction strength has been changed by a ratio of $J'/J=2$, so that while the classical ray caustic remains fixed ($C_1=\pm \sqrt{4C_2^{3}/27}$, red dashed line), the interference fringes of the wavefunction scale according to the Berry indices in the directions indicated.    \textbf{Panel (c):} $\vert \Psi \vert^2$ for the TFIM (blue shading truncated at 0.2 for clarity) is enclosed by the light cone (black, dashed). Black dots mark the locations of vortices (see section \ref{sec:Vortices}). \textbf{Inset:} Slice along the time axis at $x/a=5$. The local structure of the exact wavefunction, Eq.\ \eqref{eq:PsiLatticeSum} (blue, solid) near the light cone is well-captured by the Airy-like representation of the wavefunction Eq.\ \eqref{eq:Airylike} (orange, dashed). Away from the caustic the Airy function approximation gradually moves out of the phase with the exact result. This is because we have expanded the generating function about the caustic and can be corrected by performing a uniform approximation. }
	\label{fig:PearceyScaling}
\end{figure*}

Dividing $\Phi(k,x,t)$ as given in Eq.\ \eqref{eq:Action} by $t$ we can identify three control parameters: $(x/t,\gamma,g)$  [we reserve the energy scale $J$ to play the role of $k$ in Eq.\ (\ref{eq:generalwavecatastrophe})]. However, rays propagate in the 2D $(x,t)$ plane rather than the full 3D control space and thus for generic values of the control parameters catastrophe theory predicts we should see folds and cusps. In fact, we find a double cone made of a cusp enclosed by two folds as shown in Fig.\ \ref{fig:DoubleCone} (double cones occur both in spin systems and in coupled 1D gases \cite{Happola2012,Langen2018}). 

Mathematically speaking, the double cone arises because Eq.\ \eqref{eq:Action} has up to four stationary points within the first Brillouin zone, as shown by the green dots in the five overlays plotted in Fig.\ \ref{fig:DoubleCone}(b).  Near the origin in Fig.\ \ref{fig:DoubleCone} all four stationary points are present, but three are quasi-degenerate so  $\Psi$ is locally dominated by a Pearcey-like function which gives the inner cone. As we cross the edges of the inner cone two stationary points annihilate (indicated by red stars in the overlays) leaving two rays which in turn annihilate at the edges of the outer cone so that locally it is dominated by an Airy function. Furthermore, the XY model has a QCP at $g=1-\gamma^2$; as the critical regime is approached the inner cone narrows and eventually collapses because the three inner stationary points in the generating function coalesce at this value of $g$. In the case of the TFIM ($\gamma=1$) \cite{Calabrese2005,Calabrese2011,Bucciantini2014}, $\Phi$ has only two stationary points and one finds a single cone with edges that are dressed by Airy functions. The insight from catastrophe theory is that the single cone is non-generic and only occurs due to the special symmetry of the Hamiltonian when $\gamma$=1.

Due to the presence of four stationary points, the careful reader might expect the XY model to show signatures of the swallowtail catastrophe. Indeed, this would generically be true, however it can be verified that the quadruple root coalescence do not occur for real $k$. It is the periodic dispersion relation of the model which keeps us from physically probing the highly-singular swallowtail point. The cusp and fold catastrophes that we observe here are however inherited from the part of the swallowtail which is physically permitted.

\section{Airy and Pearcey Functions}
\label{sec:pearcey_airy_fns}

Let us now demonstrate explicitly how the Airy and Pearcey catastrophe integrals emerge in the CA. Starting with the Pearcey integral, consider first the triple stationary point coalescence responsible for the inner cone, which we have isolated in  Fig.\ \ref{fig:DoubleCone}(c). One obvious difference between this wavefunction and the Pearcey function shown in  Fig.\ \ref{fig:PearceyAiry} is that  the cone boundary in the former  is straight rather than the standard curved form of the cusp $C_{1}=\pm \sqrt{4 C_{2}^3/27}$.  Physically, this is due to the free propagation of the fermionic quasiparticles. The required transformation to take us between physical coordinates  and those of the standard curved cusp is similar to that used by Kaminski and Paris in Ref.\ \cite{Kaminski1999}. In Appendix \ref{appdx:causticsXY} we show that for our spin model  it is
\begin{eqnarray}
C_1 & = & -\sqrt{2}x/[v_{\text{\tiny{I}}}\left(t\Gamma\right)^{\frac{1}{4}}] \\
C_2 & = & -\sqrt{t}(\gamma^2+g-1)/[\sqrt{\Gamma}(g-1)], 
\end{eqnarray}
where $\Gamma=\frac{(g^3-1-2\gamma^2+3\gamma^4+g(3-2\gamma^2)+g^2(4\gamma^2-3))}{12(g-1)^3}$ and we
 have defined the \textit{Ising velocity},
\begin{equation}\label{eq:vIsing}
v_{\text{\tiny{I}}}\equiv \begin{cases}
\frac{2Jag}{\hbar}&0<|g|<1\\
\frac{2Ja}{\hbar} &1<|g|
\end{cases}\;,
\end{equation}
which is equal to $v_{\text{\tiny{LR}}}$ in the TFIM limit (in principle, $v_{\text{\tiny{LR}}}$ can be analytically solved for in closed form for general $\gamma$, however, the expression is complicated, and little physical insight is gained from writing it here). 

To complete the diffraction integral we also need the integration variable $s$. This reads $s=\sqrt{2}a\left(t\Gamma\right)^{\frac{1}{4}}k$ and results in the Pearcey-like wavefunction $\Psi_{\mathrm{Pe}}(C_1,C_2;J)$ written out in Eq.\ \eqref{eq:PearceyLike}. It rapidly tends to a true Pearcey function at longer times when $S=\sqrt{2}\pi\left(t\Gamma\right)^{\frac{1}{4}} \gg 1$.

\begin{widetext}
	\begin{eqnarray}\label{eq:PearceyLike}
		\Psi_{\mathrm{Pe}}(C_1,C_2;J) & \approx & \frac{1}{2\pi}\left(\frac{J(\gamma^2+g-1)}{\hbar v_{\text{\tiny{I}}}(g-1)C_2}\right)^{\frac{1}{2}}\int_{-S}^{S}\mathrm{d}s\;\mathrm{e}^{-\frac{\mathrm{i}J}{\hbar}\Phi_2(s;C_1,C_2)}\stackrel{Jt/\hbar\gg 1}{\propto}\left(\frac{J}{\hbar}\right)^{\frac{1}{4}}\mathrm{Pe}\left[\left(\frac{J}{\hbar}\right)^{\frac{3}{4}}C_1,\left(\frac{J}{\hbar}\right)^{\frac{1}{2}}C_2\right] \\
	\label{eq:Airylike}
		\Psi^{\mathrm{\gamma=1}}_{\mathrm{Ai}}(C^j;J) & \approx & \frac{1}{2\pi t^{1/3}}\left(\frac{2Jg^{\frac{2-j}{3}}}{v_{\text{\tiny{I}}}\hbar}\right)^{\frac{1}{2}}\int_{s_j^{\text{\tiny{Min}}}}^{s_j^{\text{\tiny{Max}}}}\mathrm{d}s_j\;\mathrm{e}^{\frac{\mathrm{i}J}{\hbar}\Phi_{1}(s_j;C^j)}\stackrel{Jt/\hbar\gg 1}{\propto}\left(\frac{J}{\hbar}\right)^{\frac{1}{6}}\mathrm{Ai}\left[\left(\frac{J}{\hbar}\right)^{\frac{2}{3}} C^{j} \right]
	\end{eqnarray}
\end{widetext}

In order to display the close resemblance between $\Psi_{\mathrm{Pe}}$ and the Pearcey function, we have plotted in Fig.\ \ref{fig:PearceyScaling} the wavefunction of the inner cone from Eq.\ \eqref{eq:PsiLatticeSum} without expansions or approximations in terms of the transformed coordinates $C_1$ and $C_2$. This can be compared with the actual Pearcey function plotted in Fig.\ \ref{fig:PearceyAiry}. The only significant deviation is near $C_2=0$. Since the limit of integration $S$ tends to 0 as $t\to 0$, the cusp point itself becomes poorly defined, and we get a `smearing' of the wavefunction as $C_2\to 0$. As a consequence, we cannot get a Pearcey function exactly at the origin, since the initial boundary condition requires the real-space wavefunction be entirely localized here. As we move away from the cusp point, however, the Pearcey function is indeed an excellent approximation to the true wavefunction.

As $C_{2}$ increases the Pearcey function can be approximated by two back-to-back Airy functions as the cusp evolves into two fold lines. Indeed, it is a general property of catastrophes that the higher ones evolve into the lower ones as we move away from the former's most singular points. This provides a rigorous explanation for why Airy functions, which are the simplest of the hierarchy of wave catastrophes, are commonly encountered in the asymptotics of light cones  \cite{Barmettler12,Perfetto17,Kormos17,Najafi2018,Eisler2013,Viti2016,Eisler2018,Hunyadi2004,Allegra2016}. 

To examine how the Airy function emerges in the CA we specialize to $\gamma=1$ (TFIM Hamiltonian). We stress that the choice of $\gamma$ does not affect the presence of the fold catastrophe (and thus Airy functions), only the simplicity of the subsequent calculations. To this end, note that for any $g\neq1$ it can be readily checked that $\Phi(\gamma=1)$ in Eq.\ \eqref{eq:Action} has only two stationary points as a function of $k$. We can therefore map onto the canonical fold generating function $\Phi_{1}(s;C)$ by expanding $\Phi$ to third order in $s$. In the CA, and up to a global phase, we show in Appendix \ref{Appdx:TFIMCaustics} that the correct control parameter in this case is
\begin{equation}
C^j(x,t)=2(x/v_{\text{\tiny{I}}}-t)(g^{2-j}/\sqrt{t})^{2/3}.
\label{eq:ControlAiry}
\end{equation}
The index $j\in\{1,2\}$ refers to cases $g>1$, and $g<1$, corresponding to above and below the QCP, respectively. The integration variable $s_j=(g^{2-j}t)^{\frac{1}{3}}(ka-\arccos[g^{3-2j}])$, and integration limits $s_j^{\text{\tiny{Min}}}=-(g^{2-j}t)^{1/3}(\pi+\arccos[g^{3-2j}])$ and $s_j^{\text{\tiny{Max}}}=(g^{2-j}t)^{1/3}(\pi-\arccos[g^{3-2j}])$ are also derived in Appendix \ref{Appdx:TFIMCaustics}.   The resulting wavefunction $\Psi^{\mathrm{\gamma=1}}_{\mathrm{Ai}}(C^j;J) $ is given in Eq.\ \eqref{eq:Airylike}. 

When $\gamma\neq 1$ this process may be repeated around each fold catastrophe, including for any inner cones, and will result in the emergence of Airy functions with different definitions of the control parameter, $C$. For example, a particular limit of Eq.\ \eqref{eq:Airylike} has been conjectured to give a universal form for the wavefront of out-of-time-ordered correlators (OTOCs) \cite{Swingle2018,Lin2018,Riddell2019a,Riddell2019b}. According to catastrophe theory this is no surprise. Furthermore, closer to the `brightest' parts of the OTOC the hierarchy of catastrophes  allows for more elaborate structures beyond the Airy function.

\begin{figure}[h!]
	\hspace{-0.2cm}\includegraphics[width=0.51\columnwidth]{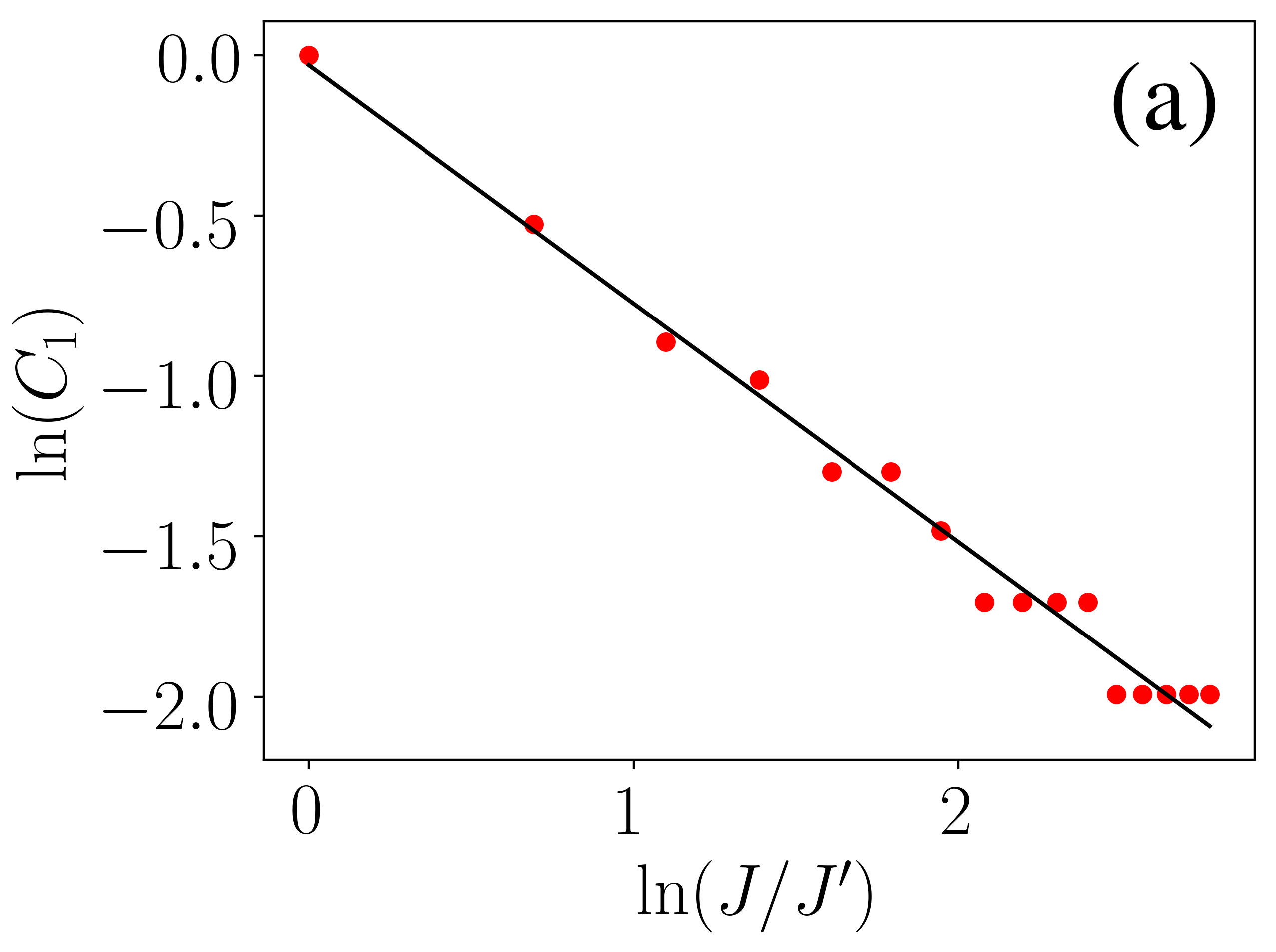}\hspace{-0.12cm}	\includegraphics[width=0.51\columnwidth]{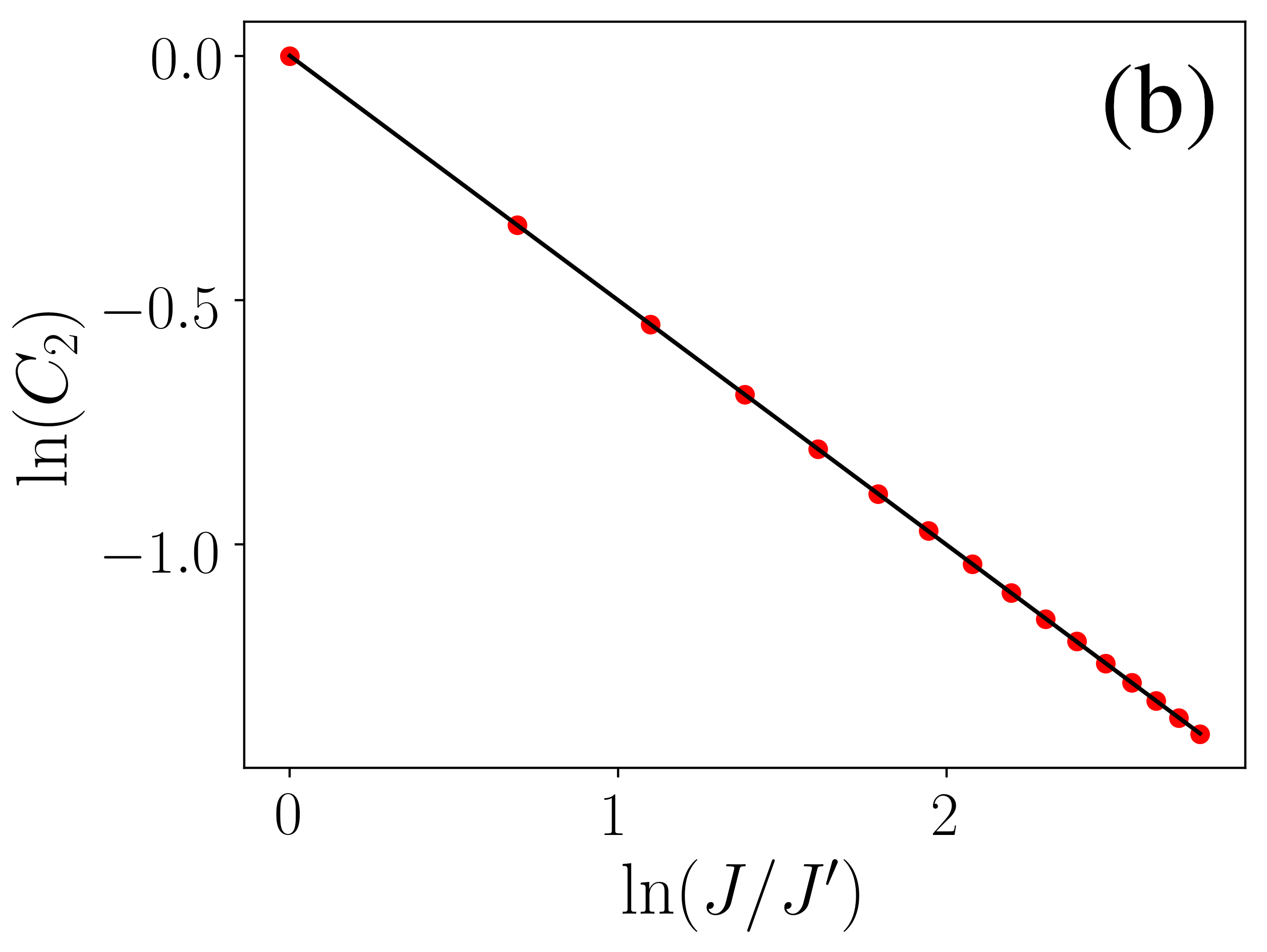}\\
	\includegraphics[width=0.5\columnwidth]{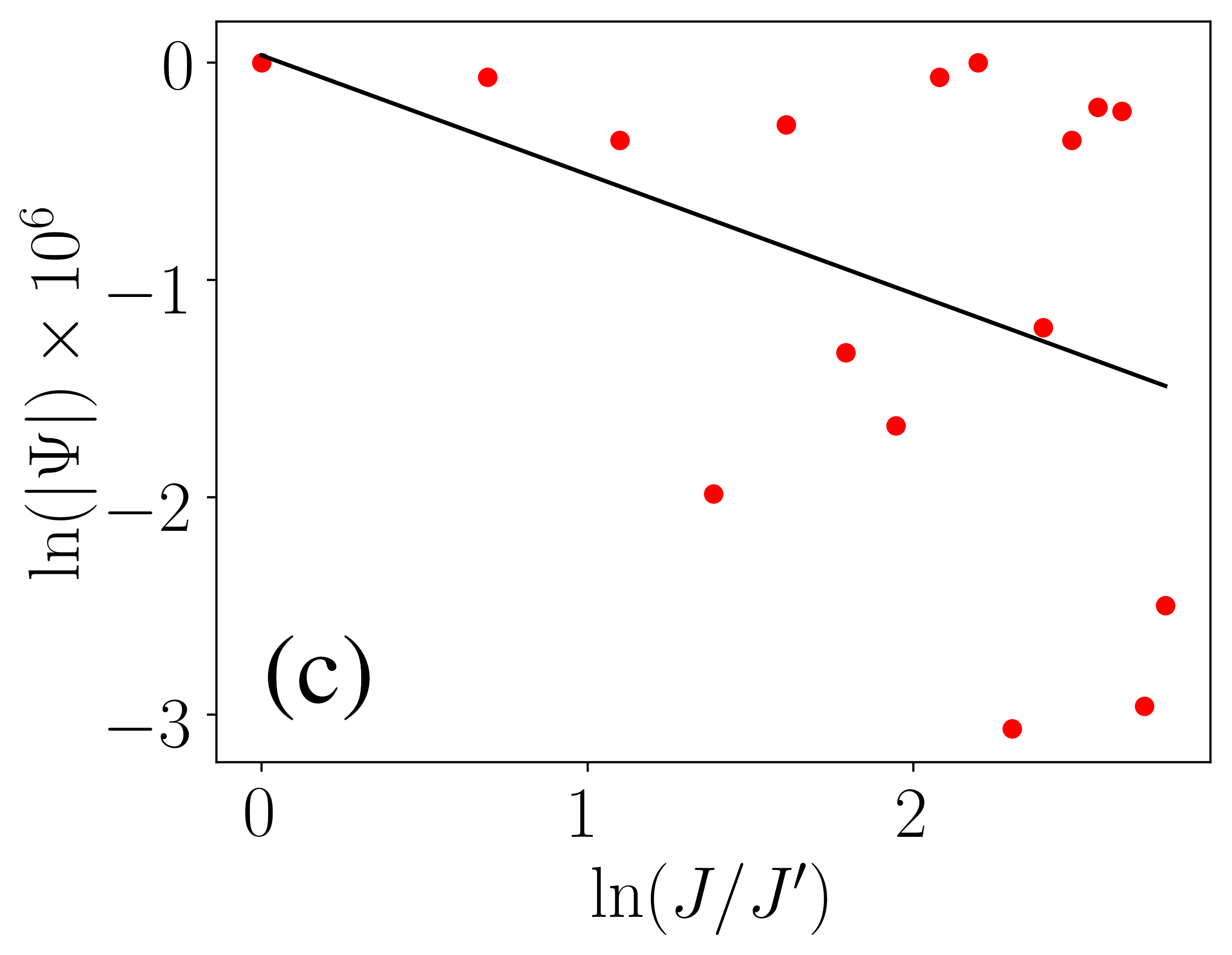}\hspace{-0.1cm}
	\includegraphics[width=0.5\columnwidth]{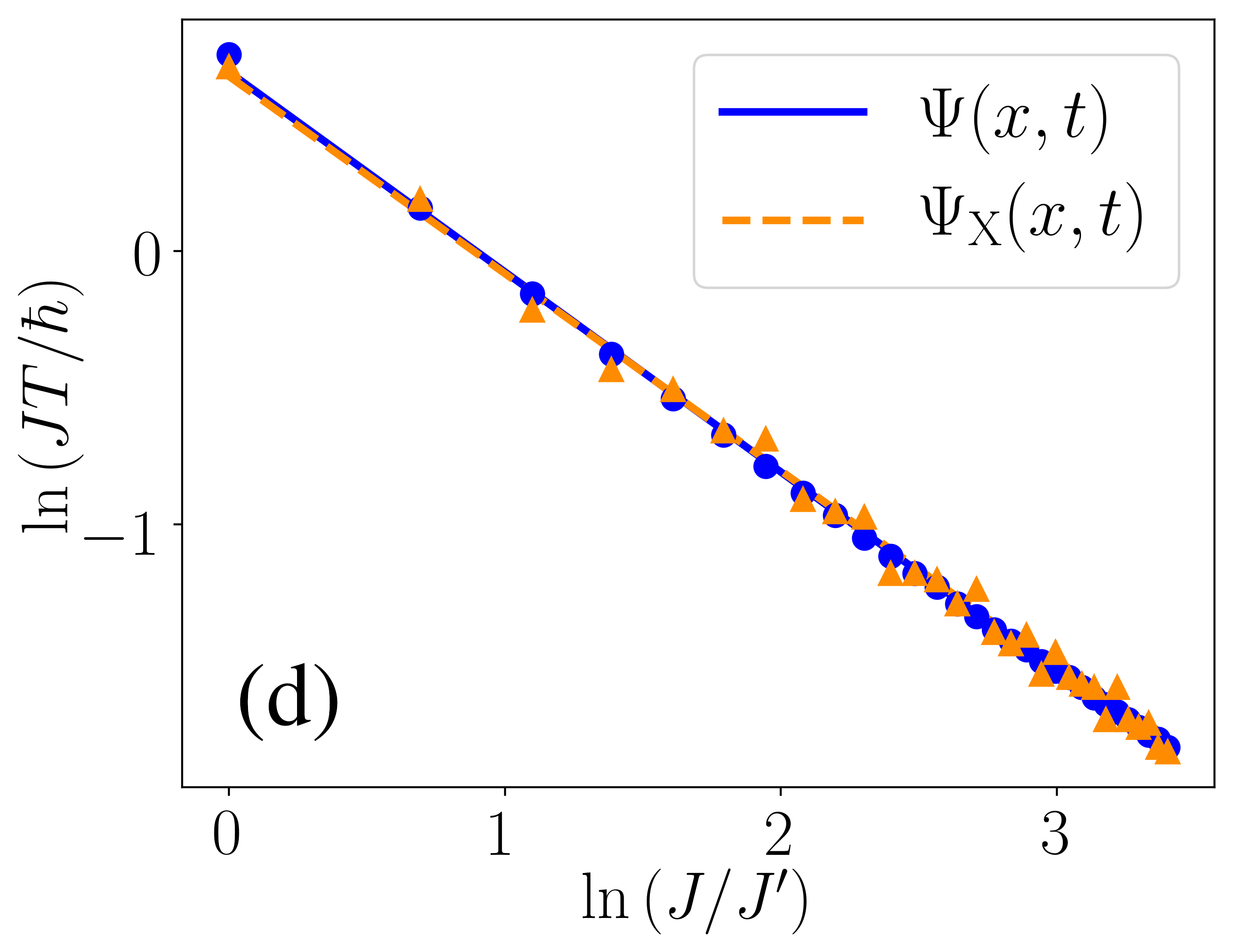}
	\caption{Self-similar scaling of light cone wavefunctions. \textbf{Panel (a):} Fringe spacing in the $C_{1}$ direction within the inner cone of the XY model scales as $J^{\varsigma_1}$ with a Berry index $\varsigma_1\approx0.743\pm0.002$ (a range of $1\leq J/J'\leq 16$ was used). The staircase pattern is due to the discreteness of the lattice. \textbf{Panel (b):} Fringe spacing scaling in the $C_2$ direction of the XY model gives a Berry index of $\varsigma_2\approx0.500\pm0.001$. \textbf{Panel (c):} Wavefunction amplitude scaling of $\ln|\Psi|\approx(-6\times10^{-7}\pm4\times10^{-7})\ln(J)$, indicating that the amplitude near the cusp has an incredibly weak scaling with $J$. This effect is explained by Eq.\ \eqref{eq:ArnoldCancellation}, since the initial condition precisely cancels the Arnol'd scaling to preserve particle number. \textbf{Panel (d):} The oscillation period, $T$, of $\Psi(x_{n},t)$ for site $x/a=5$ in the TFIM with $g=3$; Eqns.\ \eqref{eq:PsiLatticeSum} and \eqref{eq:XState} are plotted in blue circles and orange triangles, with blue-solid and orange-dashed trendlines, respectively (a range of $1 \leq J/J' \leq 30$ was chosen). Accounting for a geometric factor of $\sin[\arctan(20)]$, we find the Berry index to be $0.654\pm0.003$ and $0.646\pm0.009$ for $\Psi$ and $\Psi_\mathrm{X}$, respectively.}
	\label{fig:IndexScaling}
\end{figure}

\section{Scaling}
\label{sec:scaling}

The way the spin coupling strength $J$ and the control parameters $\mathbf{C}$ appear in combination on the right hand sides of Eqns.\ (\ref{eq:PearceyLike}) and (\ref{eq:Airylike}) shows that light cones have nontrivial scaling properties: varying $J$ is equivalent to rescaling the amplitude and coordinates.  More specifically, increasing $J$ causes the amplitude to increase at a rate determined by the Arnol'd index, and the interference patterns to oscillate more quickly in space and time at rates determined by the Berry index for each particular direction. The overall picture is that the fringes flow in towards the origin as $J$ is increased and in the (singular) classical limit, which occurs when $J \to \infty$, all wave structure is pulled into the origin.  There are other choices we could have made for the scaling parameter since it need only fill the role of $\lambda$ in Eq.\ \eqref{eq:generalwavecatastrophe}: for the TFIM we could have alternatively chosen $a$ or $g$, and in the case of the XY model we could also have chosen either of these or even $\gamma$. It is usually necessary to keep some physics constant during the scaling: we can keep the position of the classical ray caustics constant as $J$ is varied  by tuning $a$ or $g$ to keep $v_{\text{\tiny{I}}}$ unchanged.

Numerical verification of the catastrophe theory predictions for both the Arnol'd and Berry indices for the exact wavefunction Eq.\ \eqref{eq:PsiLatticeSum} is presented in Fig.\ \ref{fig:IndexScaling}. Panels (a)-(c) show the scaling in the inner cone of the XY model: the fringe scaling is obtained by measuring the distance between peaks of the wavefunction along coordinates $C_1$ and $C_2$ as $J$ is varied and match the Pearcey scaling given in Table \ref{tab:catastrophetable} to within 1\%. At first glance, it appears that panel (c) shows a contradiction between the expected amplitude scaling of the catastrophe integral and the wavefunction. However, a quick calculation involving the prefactor of the wavefunction which ensures that particle number is conserved shows that
\begin{equation}\label{eq:ArnoldCancellation}
	C_2^{-\frac{1}{2}}\sim \left(\frac{J}{J'}\right)^{-\frac{1}{4}}\;,
\end{equation}
which exactly cancels the Arnol'd scaling. This is a peculiarity of our non-generic initial condition of starting with a completely localized initial state: when tracking a particular fringe, it will move towards the origin but this normalization factor means that its height does not scale with $J$.

Panel (d) of Fig.\ \ref{fig:IndexScaling} shows the predictions in the TFIM for the period $T$ of oscillations near the caustic. Data is shown both for the exact wavefunction,  given in Eq.\ \eqref{eq:PsiLatticeSum}, and also the  `spin-flip'  state $\Psi_{\text{X}}$, given in \eqref{eq:XState}, which is easier to realize experimentally. Since the Berry index $\varsigma$ for the fold defines scaling perpendicular to the caustic, a geometric factor dependent on $v_{\text{\tiny{LR}}}$ must be applied. Numerical agreement to within 3\% of Airy scaling given in Table \ref{tab:catastrophetable} is found in both cases even for finite-sized systems at finite times.

\begin{figure}[t!]
	\includegraphics[width = 0.49\columnwidth]{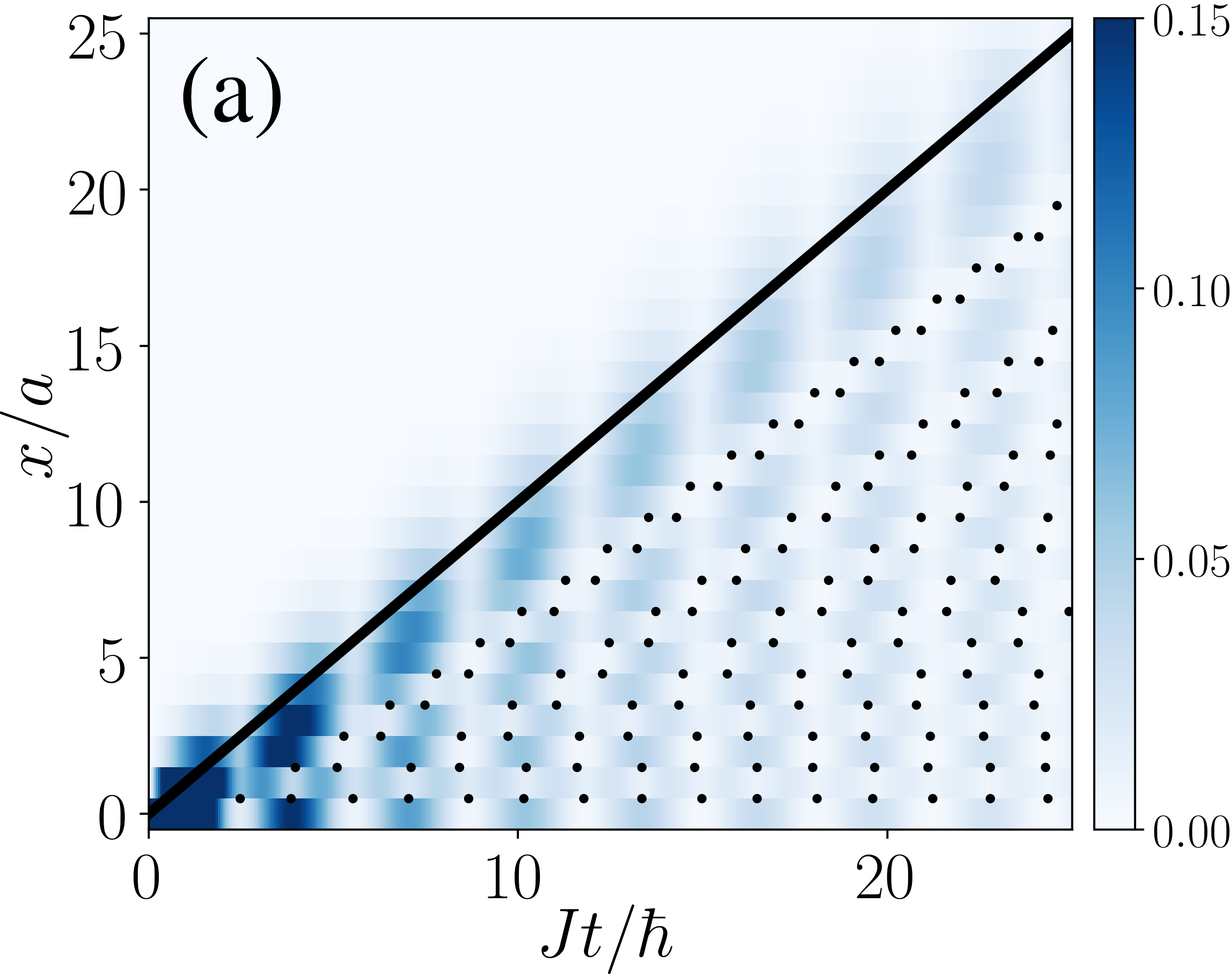}
	\includegraphics[width = 0.49\columnwidth]{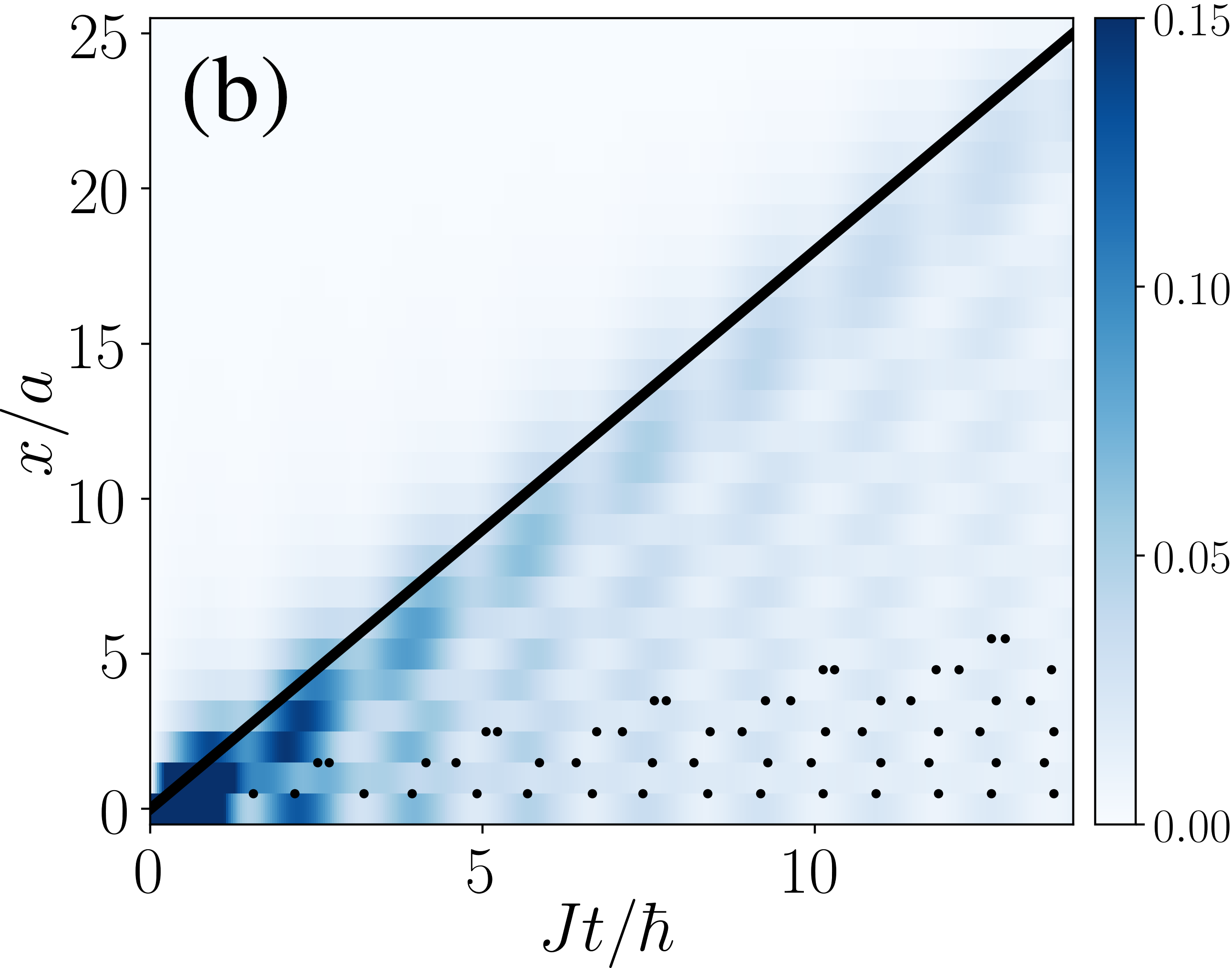}	
	\caption{\textbf{Panel (a):} Correlation function amplitude $|G(0,x_m,t)|$ for $\gamma=1$, and $g=0.5$, where the blue shading has been truncated at 0.15 for clarity.  Black dots indicate vortices. \textbf{Panel (b):} Same correlation function as in (a), now with $g=0.9$. Note that the number of vortices within the cone decreases drastically near the critical point (see Section \ref{sec:Vortices}). }\label{fig:Fig5}
\end{figure}

\section{\label{sec:Correlation}Correlation Functions and Higher Order Catastrophes}

Rather than the probability distribution associated with the wavefunction itself, light cones are usually observed in correlation functions \cite{Cheneau2012,Fukuhara2013,Langen13,Richerme14,Jurcevic2014}.  The equal time site-site correlation function is defined as 
\begin{equation}
G(x_n,x_m,t)=\braket{b^\dagger_{n} b_{m}(t)}-\braket{b^\dagger_n(t)}\braket{b_{m}(t)} \ .
\end{equation}
 Because Bogoliubov fermions are conserved, $\braket{b^\dagger_n(t)}=\braket{b_{m}(t)}=0$, and the last term vanishes.  The remaining piece is
\begin{align}
\braket{b^\dagger_{n} b_{m}(t)}=&\;\braket{\Psi(t)|b^\dagger_{n} b_{m}|\Psi(t)}\nonumber\\=&\;\frac{1}{N}\sum\limits_{k,k'}\mathrm{e}^{-\mathrm{i}(\epsilon_k-\epsilon_{k'})t/\hbar}\prescript{}{b}{\bra{0}}\tilde{b}_{k'}b^\dagger_{n} b_{m}\tilde{b}_k^\dagger\ket{0}_b\;.
\end{align}
where we have used the state vector $\vert \Psi(t) \rangle$ given in Eq.\ \eqref{eq:PsiVector2}.  Expressing all the operators in terms of quasimomentum (see Appendix \ref{Appdx:BogoliubovDynamics}) we obtain
\begin{align}\label{eq:Correlation}
G(x_n,x_m,t)=&\;\frac{1}{N^2}\sum\limits_{k,k'}\mathrm{e}^{-\mathrm{i}(\epsilon_k-\epsilon_{k'})t/\hbar}\mathrm{e}^{\mathrm{i}(kx_m-k'x_n)}\nonumber\\=&\;\Psi(x_m,t)\Psi(-x_n,-t)\;.
\end{align}
In Fig.\ \ref{fig:Fig5} we plot $G(0,x_m,t)$ on the upper half of the spin chain for two different values of $g$. It displays the same features as the wavefunction: a light cone, interference fringes, and vortices.
In the CA, the equal time site-site correlation function becomes
\begin{align}
G_{\mathrm{CA}}(x,x',t)=&\;\frac{a}{(2\pi)^2}\int_{-\pi/a}^{\pi/a}\int_{-\pi/a}^{\pi/a}\mathrm{d}k\,\mathrm{d}k'\;\mathrm{e}^{\mathrm{i}\left(\Phi(k,x)-\Phi(k',x')\right)}\nonumber\\=&\;\Psi_{\mathrm{CA}}(x,t)\Psi_{\mathrm{CA}}(-x',-t)  \ ,
\end{align}
and expanding around the cone boundaries gives
\begin{equation}
G_{\mathrm{CA}}(x,x',t)\approx \Psi_{\mathrm{Ai}}(C(x,t),t)\Psi_{\mathrm{Ai}}(C(-x',-t),-t) \ ,
\end{equation}
where $C(x,t)$ is the same function of $x$ and $t$ as that given in Eq.\ (\ref{eq:ControlAiry}).

Measurements and calculations (based on doublon and holon quasiparticles) on the BH model following a quench also find a product of two Airy functions for $G(x_n,x_m,t)$ \cite{Barmettler12,Cheneau2012}. However, referring to Table \ref{tab:catastrophetable}, generic dimension 3 singularities (i.e.\ two spatial coordinates $x_{n}$ and $x_{m}$, as well as time $t$) of corank 2 (i.e.\ 2 integration variables, like in the two-site correlation function) are the elliptic umbilic, and hyperbolic umbilic catastrophes. The elliptic umbilic diffraction catastrophe has been studied by Berry, Nye and Wright \cite{Berry1979} via the optics of a triangular water droplet lens, while the hyperbolic umbilic is a direct consequence of the primary coma aberration \cite{Berry1980a}, and has been observed in matter waves using electron microscopy \cite{Petersen2013}. 
These catastrophes are generally more complicated than a squared Airy function, however, we note that in a certain plane the hyperbolic umbilic wave catastrophe does indeed reduce to the product of two Airy functions. More precisely, the hyperbolic umbilic wave catastrophe is given by \cite{NIST}
\begin{equation}
	\Psi_{\mathrm{HU}}(x,y,z)= \lambda \iint \limits_{-\infty}^{+\infty} \mathrm{d}s_1\mathrm{d}s_2 \;\mathrm{e}^{\mathrm{i} \lambda \left(s_1^3+s_2^3+C_3s_1s_2+C_2s_2+C_1s_1\right)}  \ ,
	\label{eq:hyper}
\end{equation}
and when $C_3=0$ this reduces exactly to 
\begin{equation}
\Psi_\mathrm{HU}(C_1,C_2,0)=\frac{4 \pi^2 \lambda^{\frac{1}{3}}}{3^{\frac{2}{3}}} \text{Ai}\left(\frac{C_1 \lambda^{\frac{2}{3}}}{3^{\frac{1}{3}} }\right)\text{Ai}\left(\frac{C_2 \lambda^{\frac{2}{3}}}{3^{\frac{1}{3}} }\right).
\end{equation}
 Thus, both the XY model and the BH model give rise to a non-generic special case.

What physical quantity could the $C_{3}$ control parameter represent? Studying the form of
$\Psi_{\mathrm{HU}}$ given in Eq.\ (\ref{eq:hyper}) we note that $C_{3}$ controls the coupling between the $s_1$ and $s_2$ variables  which in a spin chain correspond to the two quasimomenta  $k$ and $k'$. For noninteracting quasiparticles, which is the case for the exactly solvable models considered in this paper, the two quasimomenta are uncoupled and thus $C_{3}$ is zero. Furthermore, the particular regime of the BH model where Refs.\ \cite{Barmettler12,Cheneau2012} obtained a product of Airy functions also corresponds to the free quasiparticle case.  It is therefore clear that $C_{3}$ can be used to parameterize quasiparticle-quasiparticle scattering, and we predict that a model with interacting quasiparticles will give rise to light cones that sample hyperbolic umbilic wave catastrophes. This feature could be verified in an experiment where the strength of the coupling is varied for then the scaling along $C_3$ should go as $\varsigma_{3}=1/3$.

Other quantities, for example the spin-spin correlation function, $\Sigma_{nm}=\braket{\sigma_n^x\sigma_m^x}-\braket{\sigma_n^x}\braket{\sigma_m^x}$, may also be calculated exactly via the Jordan-Wigner and Bogoliubov transformations, and simplified using Wick's theorem. The functional forms of these quantities in the continuum approximation remain diffraction integrals, and thus will also display universal behaviour corresponding to catastrophes.

\begin{figure}[t]\centering
	\includegraphics[width=0.7\columnwidth]{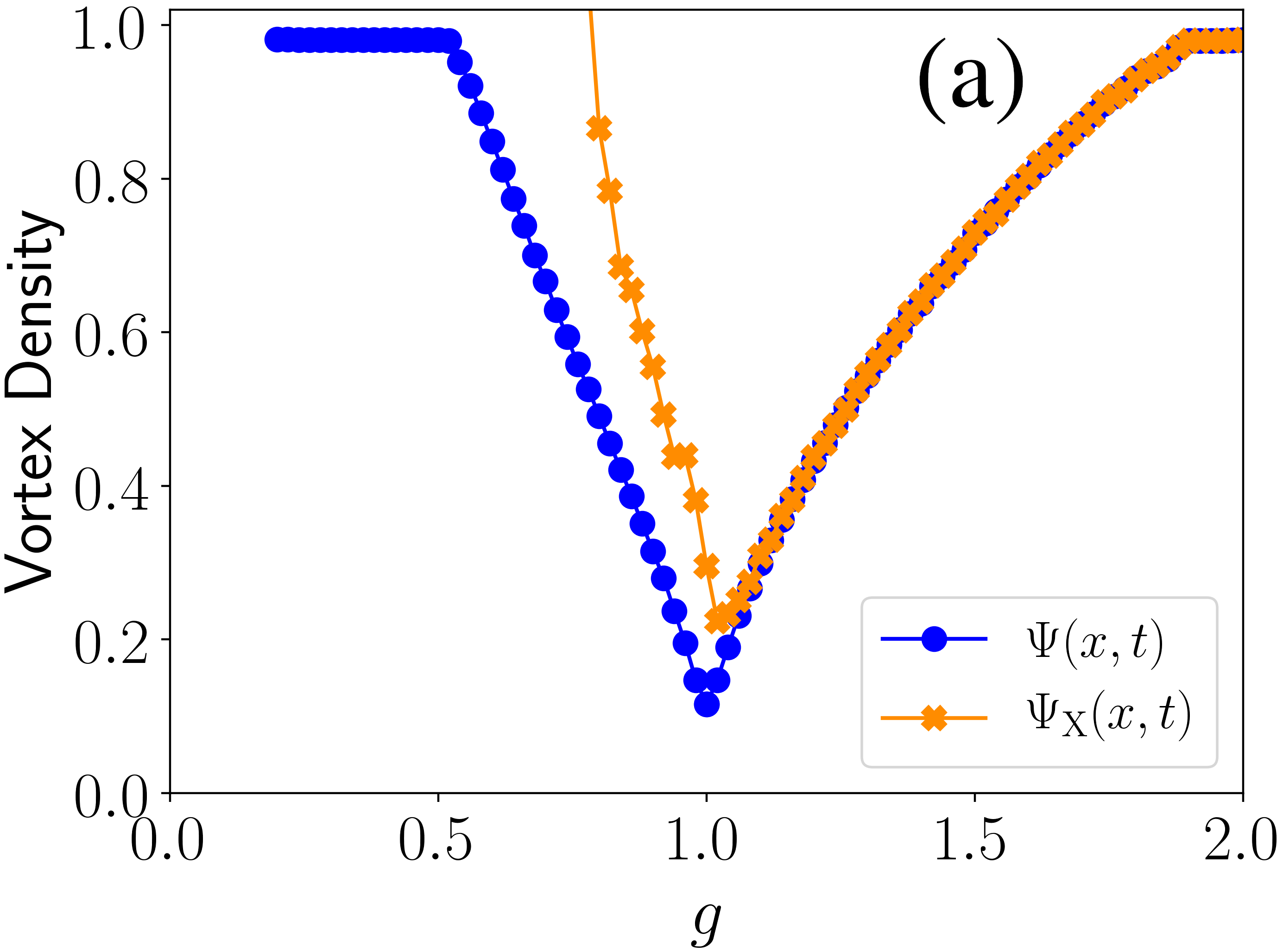}\\
	\hspace{-2mm}\includegraphics[width=0.7\columnwidth]{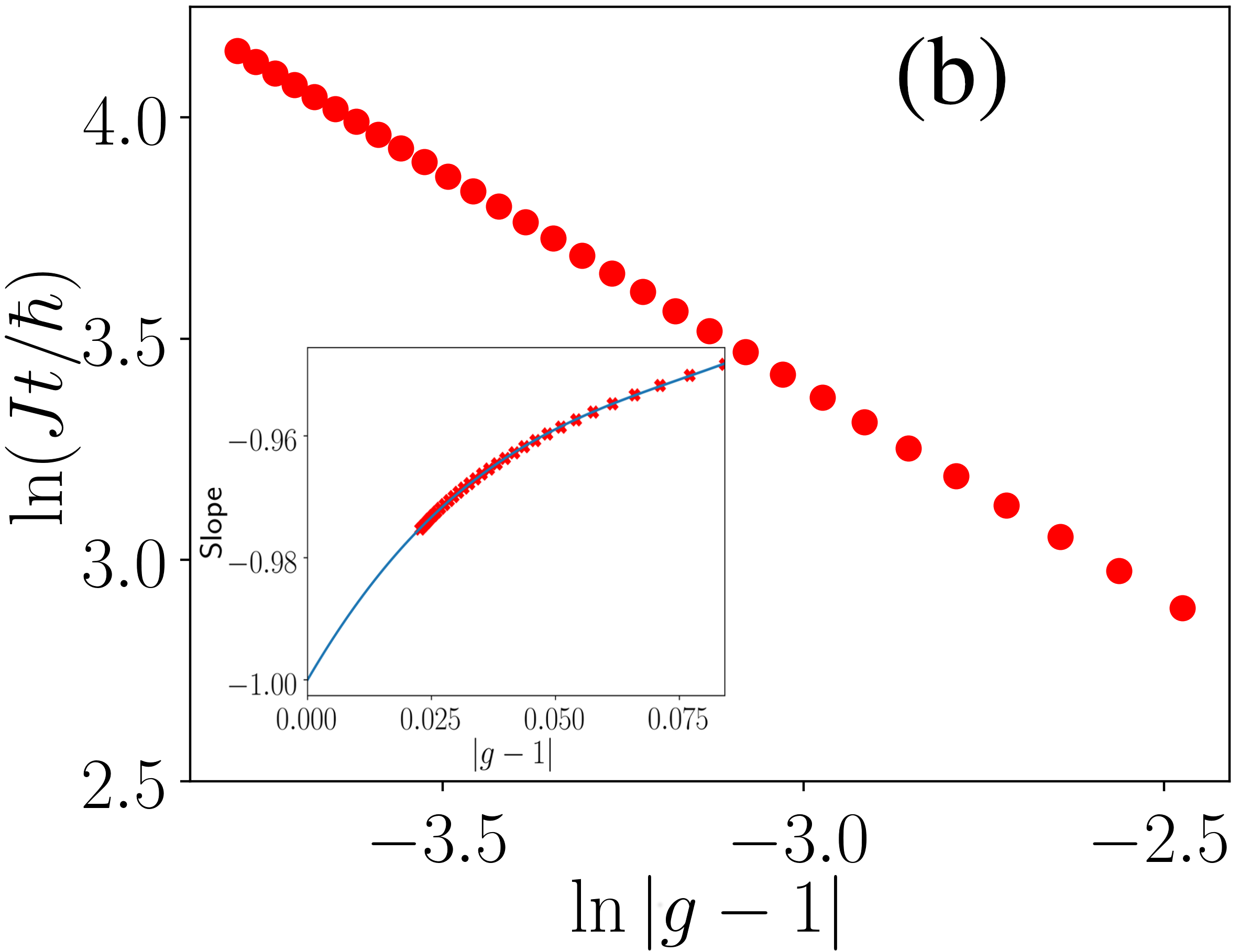}
	\caption{\textbf{Panel (a):} Vortex density inside the TFIM light cone reaches a sharp minimum at the QCP for both Eq.\ \eqref{eq:PsiLatticeSum} and the spin-flip state Eq.\ \eqref{eq:XState}. We define vortex density as being the total number of vortices that occur within a cone up to the time at which the light cone hits the edge of the system, taking care to normalize for different cone sizes at different values of $g$.   \textbf{Panel (b):} Numerical determination of vortex pair creation times at a fixed point in space as $g=g_c=1$ is approached. In order to extrapolate to the critical point (inset), 30 data points ($g$,$Jt/\hbar$) are fitted to a quadratic and then differentiated. The resulting slopes are extrapolated to $g_c$  using a cubic and the intercept gives  $\nu z=0.9999\pm 0.0004$ (standard error on the fit). The range $0.02\leq |g-1|\leq0.12$ of $g$ was chosen to optimize the proximity to the critical point along with data accuracy, since the wavefunction becomes highly oscillatory as $g\to 1$. Numerical errors are smaller than the symbol sizes.}\label{fig:Fig6}
\end{figure}

\section{\label{sec:Vortices}Vortices and Criticality}

As seen in Figs.\  \ref{fig:PearceyScaling} and \ref{fig:Fig5}, and also Fig.\ \ref{fig:LightCones2} in the appendices, we find that light cones contain lattices of vortex-antivortex pairs. Vortices form the fine structure of wave catastrophes \cite{Berry1980a,Nye1974,Kaminski1999,Nye2006}, and in a continuum are zeros of $\Psi$ where the phase $\chi\equiv\text{Arg}\Psi$ is undefined (takes all values) and has the topological property
\begin{equation}
	\oint_\mathcal{C}\mathrm{d}\chi=\pm 2\pi\;,
\end{equation}
where $\mathcal{C}$ is any closed path which contains a single vortex.  On a discrete lattice we can still use such circuits to find vortices, but across lattice sites one must perform a sum instead of integrating, meaning that their spatial position is only known up to the lattice constant: in figures we place the vortices between lattice sites. Furthermore, vortices on a lattice need not correspond to nodes or even phase singularities, but to points where the phase difference between adjacent sites is $\pm \pi$ (i.e.\ phase kinks or dark solitons). Thus, while phase interference
regulates the amplitude divergence of ray caustics, the effect of a lattice is to regulate the phase singularities of wave theory. In recent work by some of the authors \cite{Mumford2019}, the regularization of phase singularities by a lattice has been considered in Fock space.

Whereas the classical light cone changes smoothly at the QCP [see, e.g., Eq.\ (\ref{eq:vIsing})], there is a sharp minimum in the vortex density, i.e.\ many vortex-antivortex pairs annihilate, see Fig.\ \ref{fig:Fig6}. In the CA, all vortices except those closest to the central axis annihilate at the QCP, while on a discrete lattice, more off-axis vortices survive but the same trend is observed. At a fixed point in space, the time at which a vortex is first detected increases as one approaches the critical point, becoming infinite in the CA. This diverging time scale $\tau$ is related to critical slowing and suggests a connection to the dynamical critical exponent, $z$. According to the scaling hypothesis of critical phenomena
\begin{equation}
\tau \propto \xi^{z} \,
\end{equation}
 where $\xi=\vert g-g_{c} \vert^{-\nu}$ is the correlation length and $\nu$ is its equilibrium critical exponent. Fig.\ \ref{fig:Fig5}(b) plots $\tau$ as found from the wavefunction Eq.\ \eqref{eq:PsiLatticeSum}  as $g$ is tuned to the QCP.  By extrapolating the numerical data [Fig.\ \ref{fig:Fig5}(b) inset] to the critical point we obtain $\nu z = 1$ and hence recover the known critical scaling for the 1D TFIM \cite{Bunder1999,Dutta2015}. For purposes of clarity, we have only included the set of vortices which annihilate closest to the axis $x=0$. Vortices which annihilate farther off-axis also display similar trends, which can be seen in Appendix \ref{Appdx:vortexscaling}, along with further figures which help with visualization of this process.
 
 While a more complete understanding of the nature of the vortex-antivortex pairs within the light cone remains a subject of future work, we wish to highlight that their presence and scaling laws provide an interesting link between the predictions of catastrophe theory and universality (in and out of equilibrium). Due to the self-dual nature of the TFIM, qualitative behaviour for $g>1$ is identical to that of the wavefunction below the transition with $g\to1/g$ and $t\to gt$.

\section{\label{sec:expt}Experimental realization: spin flip state}

The structural stability of catastrophes explains why they occur so frequently in nature. Apart from the examples given in the Introduction, they can also occur in disordered systems such as at the Anderson transition where an evanescent Airy function occurs \cite{Lemarie2010}, and it has also been shown that wave catastrophes have the property of self-healing after being disrupted \cite{Ring2012}.  There are, therefore, a broad range of initial conditions and spin models which will give rise to caustics in their dynamics.

So far we have used the initial condition of a localized single quasiparticle, as given in Eq.\ (\ref{eq:PsiVector1}). This is a non-generic initial condition and the reader may question how generic the resulting light cones really are. In fact, all our analysis is stable to perturbations around this initial condition.  In particular, a state which is naturally generated in trapped ion experiments where individual ions can be addressed is a spin flip state which starts with all spins polarized in the $x$ direction, except for the central spin, say, which is flipped \cite{Jurcevic2014},
\begin{equation}\label{eq:XState}
	\Psi_\mathrm{X}(x,t)\equiv\bra{x}\mathrm{e}^{-\mathrm{i}Ht/\hbar}\ket{\uparrow^x \ldots \uparrow^x\downarrow^x\uparrow^x \ldots \uparrow^x}\;.
\end{equation}
It is important to realize that physical spins are in general superpositions of multiple quasiparticles and vice versa. We elaborate upon the mathematical details of this point in Appendix \ref{Appdx:SpinFlip}. What we find is that as long as the quench is not too close to the transition the number of quasiparticles created by a spin flip is close to one and hence we are perturbing around the single quasiparticle state given in Eq.\ (\ref{eq:PsiVector1}). The evidence for this statement can be found in Figs.\ \ref{fig:IndexScaling}(d) and \ref{fig:Fig6}(a), which compare the results of using $\Psi_{\text{X}}$ with those of $\Psi$.  We find that the scaling properties are essentially identical in the two cases whilst the behavior of the vortex density shows some finite differences but is qualitatively the same.

\section{Discussion and Conclusions}
\label{sec:conclusion}

Caustics are a natural phenomenon that can be seen by looking up in the sky on a rainy day. The primary bow of a rainbow is a fold caustic and careful observation reveals supernumerary arcs that are interference fringes described by the Airy function. This is the first in a hierarchy of caustics of increasing complexity whose underlying description is via catastrophe theory. This hierarchy has previously been explored in optics (particularly in the field of gravitational lensing \cite{Nye}), thermodynamics \cite{Poston96,Gilmore81}, laser physics \cite{Gilmore77,Gilmore78}, hydrodynamics \cite{Ursell1994,Berry05,Stone2017} and also cosmology \cite{Zeldovich82,Feldbrugge18}. By showing that light cones in many-body systems are also caustics, we are able to open the door to the application of a rigorous and unified mathematical framework for describing the dynamics of these systems following a quench.  

The main conceptual result of this paper is that there is a hierarchy of light cone structures. They are stable against perturbations and dressed by characteristic wavefunctions that scale according to the sets of exponents given in Table \ref{tab:catastrophetable}. The fold catastrophe and its attendant Airy function features in the TFIM, but breaking the symmetry of the TFIM leads us to the XY model and the second catastrophe,  the cusp, which is dressed by the lesser-known Pearcey function.  Choosing the spin coupling $J$ as a tuning parameter, we show how the scaling exponents lead to non-trivial scaling of these wave catastrophes as $J$ is varied.

The TFIM and XY models are exactly solvable and hence their quasiparticles are noninteracting. However, the defining feature of catastrophe theory is that it deals with structurally stable singularities and hence the light cone caustics we have described also occur in the presence of perturbations such as weak quasiparticle interactions. A related example of this is provided by the celebrated Kolmogorov-Arnold-Moser (KAM) theorem which shows that tori in the phase space of integrable systems are stable against nonintegrable perturbations. There is in fact a close connection between caustics and the quasiperiodic motion that arises in dynamical systems due to the existence of the tori \cite{Arnold97}.





Higher order catastrophes will become important in higher dimension spin lattices. Another way that higher order catastrophes become important is through $n$-body correlation functions. For the TFIM we find that the two-site equal time correlation function is described near the cone edge by the product of two Airy functions, which is, however, a special case of the hyperbolic umbilic catastrophe. We predict that adding quasiparticle interactions will lead to the full hyperbolic umbilic catastrophe. 

On their finest scales, wave catastrophes contain vortex-antivortex pairs. We have seen that in the case of light cones in 1D spin chains these become vortex-antivortex pairs in space-time. We note in passing that these are reminiscent of the Kosterlitz-Thouless transition that occurs in one space and one time dimension in the quenched 1D Bose-Hubbard model \cite{Gardas2017} and in quantum wires \cite{Zaikin1998}. Being high energy features, we find that the vortices are strongly affected by critical slowing near a QCP, unlike the light cone itself which evolves smoothly. The vortices contain all the information about the QCP and can be used to extract the critical scaling behavior.
 
 The fact that light cones are structurally stable and fall into distinct classes, each of which has its own set of scaling exponents, underlines that as a phenomenon they are an example of universality in out-of-equilibrium dynamics, somewhat akin to the universality classes of equilibrium phase transitions. The underlying reason for this universality in both cases is the presence of singularities, and the realization that light cones are caustics aids us in identifying and understanding their properties.

\acknowledgments We are grateful to Marc Cheneau for first pointing out to us the existence of Airy functions in light cones and to Laurent Sanchez-Palencia for discussions. We thank the Natural Sciences and Engineering Research Council of Canada (NSERC) for funding.

\appendix

\section{\label{Appdx:BogoliubovDynamics}Dynamics of a Bogoliubov Fermion}

The spin models dealt with in this paper can be exactly diagonalized in terms of Bogoliubov fermions. Their Hamiltonians can therefore be written in the form
\begin{equation}
H=\sum_k\epsilon_k\left(\tilde{b}_k^\dagger\tilde{b}_k-\frac{1}{2}\right) 
\label{eq:fermionHamiltonian}
\end{equation}
where $\epsilon_k$ is the dispersion relation and the operators $\tilde{b}^{\dag}_{k}$ and $\tilde{b}_{k}$ create and annihilate, respectively, fermions with quasimomentum $k$. We shall denote the action of the creation operator on the Bogoliubov vacuum as  $\tilde{b}^{\dag}_{k} \vert 0 \rangle_{b}=\vert k \rangle_{b}$. These operators are related to their counterparts in position space via a discrete Fourier transform:
\begin{eqnarray}
b_{x} & = & \frac{1}{\sqrt{N}} \sum_{k} e^{-\mathrm{i}kx} \ \tilde{b}_{k} \\
\tilde{b}_{k}  & = & \frac{1}{\sqrt{N}} \sum_{x} e^{\mathrm{i}kx} \ b_{x}  
\end{eqnarray}
where $N$ is the number of sites/spins. 

Applying the time evolution operator to a single Bogoliubov fermion created at the center of the lattice we obtain the state vector:
\begin{align}
\ket{\Psi(t)}=\;&e^{-\mathrm{i} H t/\hbar}b^\dagger_{r=0}\ket{0}_b=e^{-\mathrm{i}\hat{H}t/\hbar}\left(\frac{1}{\sqrt{N}}\sum\limits_{k}\tilde{b}_k^\dagger\right)\ket{0}_b\\
=&\;\frac{e^{\mathrm{i}\theta(t)}}{\sqrt{N}}\sum\limits_{k}e^{-\mathrm{i}\epsilon_kt/\hbar}\ket{k}_b
\end{align}
where  $\theta(t)\equiv(t/2\hbar)\sum_k\epsilon_k$. The corresponding spatial wavefunction is
\begin{equation}
\Psi(x,t)=\prescript{}{b}{\braket{x|\Psi(t)}}=\frac{e^{\mathrm{i}\theta(t)}}{\sqrt{N}}\sum\limits_{k}e^{-\mathrm{i}\epsilon_kt/\hbar}\prescript{}{b}{\braket{x|k}_b}\;,
\end{equation}
and inserting the standard result $\braket{x|k}=e^{\mathrm{i}kx}/\sqrt{N}$ for the overlap gives
\begin{equation}\label{eq:PSISum}
\Psi(x,t)=\frac{e^{\mathrm{i}\theta(t)}}{N}\sum\limits_{k=-\frac{\pi}{a}}^{\frac{\pi}{a}-\frac{2\pi}{Na}}e^{\mathrm{i}(kx-\epsilon_kt/\hbar)}\;.
\end{equation}
If we allow $\Delta k=2\pi/(aN)$ to become very small ($N>>1$) we can approximate the sum by the integral
\begin{equation}\label{eq:Integral}
\Psi(x,t)= \frac{e^{\mathrm{i}\theta(t)}\sqrt{a}}{2\pi}\int_{-\pi/a}^{\pi/a}\mathrm{d}k\;e^{\mathrm{i}\Phi}
\end{equation}
with generating function $\Phi=kx-\frac{t}{\hbar}\epsilon_{k}$. A comparison of the discrete and continuum cases for the TFIM is given in Fig.\ \ref{fig:LightCones2}. In the semiclassical regime ($N>>1$) both the sum and the integral are dominated by the points at which $\Phi$ is stationary. Along the caustic, however, a saddle-point approximation fails since we are at a degenerate stationary point.

\section{Diagonalization of the XY Model}
\label{appdx:diagonalization}

The Hamiltonian for the XY model is,
\begin{equation}
\hat{H}=-\sum_{\braket{ij}}\left(J_x\sigma_{i}^x\sigma_{j}^x+J_y\sigma_{i}^y\sigma_{j}^y\right)-h\sum_i\sigma_i^z
\end{equation}
where $\sigma_i^\alpha$, $\alpha \in \{x,y,z\}$ are the Pauli operators for the $i$th site. We will use the Jordan-Wigner (JW) transformation, followed by a Bogoliubov rotation, in order to diagonalize $H$. Following the conventions used by Dutta \textit{et al.} in Ref.\ \cite{Dutta2015}, the transformation to JW fermions is given by, 
\begin{align}
\sigma_i^z=&\;2c_i^\dagger c_i-1\label{eq:JWSz}\\
\sigma_i^-=&\;c_i\prod_{j<i}(1-2c_j^\dagger c_j)=-(c_i+c_i^\dagger)e^{\mathrm{i}\pi\sum_{j<i}c_j^\dagger c_j}\;\label{eq:JWSm}.
\end{align}
We note that the JW fermions and Bogoliubov fermions have different vacuua; some more discussion of this point can be found in Appendix \ref{Appdx:SpinFlip}.

Next, we use a Fourier transform, $\tilde{c}_k^\dagger=\sum_j\mathrm{e}^{\mathrm{i}kx_j}c_j^\dagger$, and then rotate to Bogoliubov fermions via
\begin{equation}
\tilde{b}_k^{\dagger}=v_k\tilde{c}_k+\mathrm{i}u_k\tilde{c}_{-k}^\dagger
\end{equation}
along with the corresponding destruction operator and transformations for $-k$. Here, $u_k\equiv \cos(\phi_k/2)$, $v_k\equiv \sin(\phi_k/2)$, and $\tan(\phi_k)=(J_y-J_x)\sin(ka)/((J_y+J_x)\cos(ka)+h)$, with properties $u_k=u_{-k}$, $v_k=-v_{-k}$ in order to ensure the anticommutation relations $\left\{\tilde{c}_k^\dagger,\tilde{c}_{k'}^\dagger\right\}=\left\{\tilde{c}_k,\tilde{c}_{k'}\right\}=0$ and $\left\{\tilde{c}_k^\dagger,\tilde{c}_{k'}\right\}=\delta_{kk'}$ hold. We can simplify the resulting Hamiltonian to get it in the form of Eq.\ (\ref{eq:fermionHamiltonian}) with $\epsilon_k=2\sqrt{h^2+J_x^2+J_y^2+2h(J_x+J_y)\cos(ka)+2J_xJ_y\cos(2ka)}$ being a function of the parameters $J_x$, $J_y$, $h$, and $a$.  

\begin{figure}
	\centering
	\includegraphics[width=0.8\columnwidth]{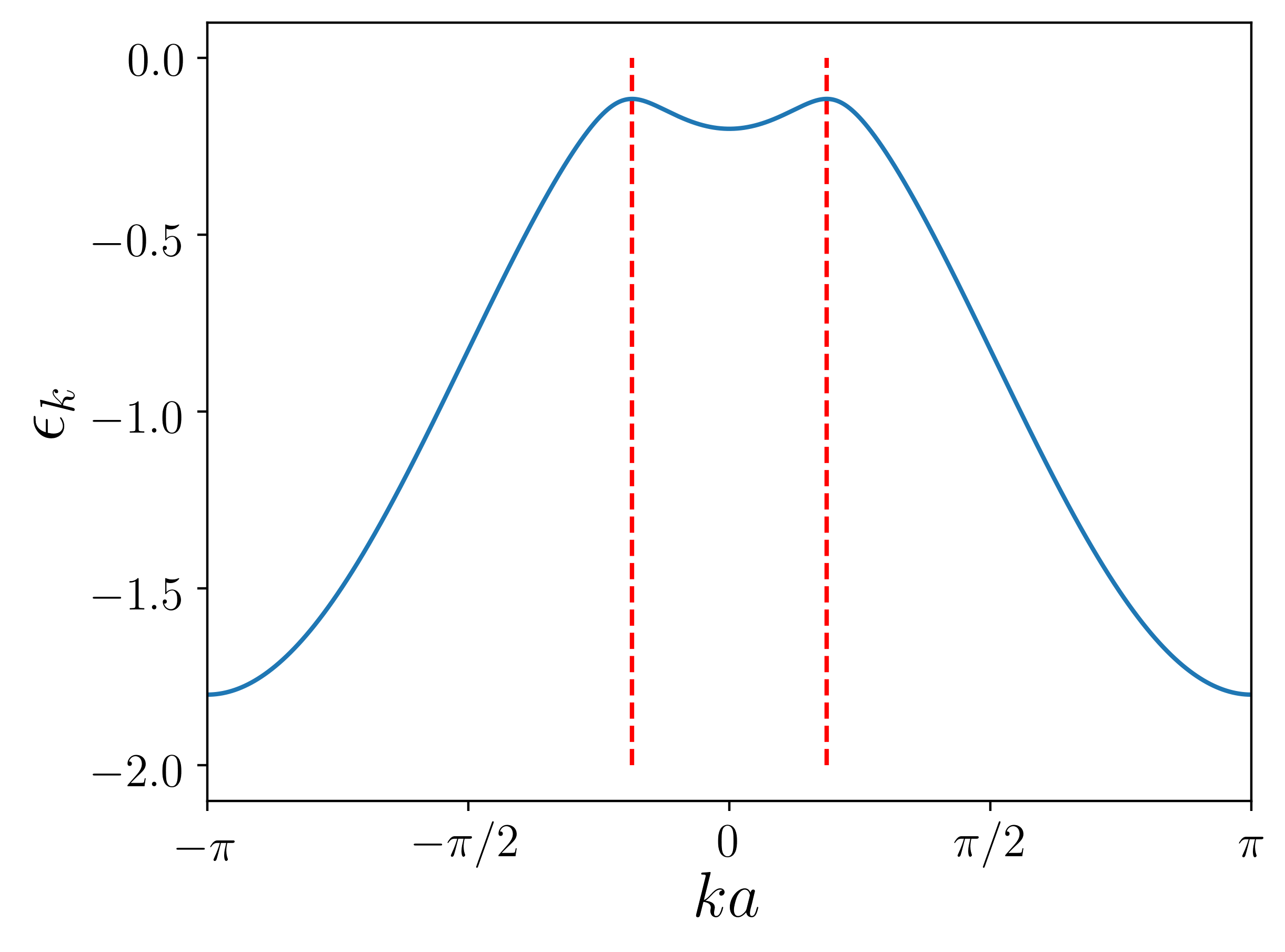}
	\caption{The XY dispersion relation, given by Eq.\ \eqref{eq:XYDispersion}, for $\gamma=0.2$ and $g=0.8$. Bearing in mind the periodicity of the dispersion relation, one can see that it has four stationary points for these parameter values. The three stationary points that are responsible for the Pearcey function are those that lie between the vertical red dashed lines. The Pearcey function is the wave catastrophe that dresses the inner cone.}
	\label{fig:XYDispersion}
\end{figure}

Next we introduce the anisotropy parameter $\gamma$ so that we can write $J_x\equiv J(1+\gamma)/2$, $J_y\equiv J(1-\gamma)/2$ and let $h\equiv gJ$. We thereby arrive at the standard form of the Hamiltonian
\begin{equation}
\frac{\hat{H}}{J}=-\frac{1}{2}\sum_{\braket{ij}}\left((1+\gamma)\sigma_{i}^x\sigma_{j}^x+\left(1-\gamma\right)\sigma_{i}^y\sigma_{j}^y\right)-g\sum\limits_i\sigma_i^z
\end{equation}
with dispersion $\epsilon_k=2J\sqrt{(\cos(ka)+g)^2+\gamma^2\sin^2(ka)}$.
If we change our conventions in order to be consistent with Sachdev \cite{Sachdev2011} we must rotate the Hamiltonian by taking $\sigma^x\to \sigma^x$, $\sigma^y\to \sigma^y$, and $\sigma^z\to-\sigma^z$.
Then we'll instead have
\begin{equation}\label{eq:XYDispersion}
	\epsilon_k=2J\sqrt{(\cos(ka)-g)^2+\gamma^2\sin^2(ka)}
\end{equation}
Effectively this is like taking $g\to-g$, allowing us to return to the standard form of the transverse-field Ising model in the $\gamma\to 1$ limit, presented in the main text. The dispersion relation given in Eq.\ (\ref{eq:XYDispersion}) is plotted in Fig.\ \ref{fig:XYDispersion}.

\section{Caustics in the XY Model}
\label{appdx:causticsXY}

In this appendix we give more details of the calculations of the caustics and their wavefunctions that are presented in the main text. The XY model contains both fold and cusp catastrophes; we focus particularly on the cusp catastrophe and defer some of the treatment of the fold catastrophe to the next appendix (Appendix \ref{Appdx:TFIMCaustics}) which is on the TFIM.

\subsection{Calculation of classical caustics in the XY model}

The light cone conditions, or equivalently the caustic conditions, are given in Eqns.\ (\ref{eq:fermat}) and (\ref{eq:causticcondition}) in the main text. These correspond to vanishing first and second derivatives of the generating function $\Phi=kx - \epsilon_{k} t/ \hbar$. The vanishing of the first derivative with respect to $k$ gives the equation
\begin{eqnarray}
	\frac{\partial\Phi}{\partial k} & = & x-  \nonumber \\ && \frac{Jt\left(2a\gamma^2\cos(ka)\sin(ka)+2a(g-\cos(ka))\sin(ka)\right)}{\hbar\sqrt{(g-\cos(ka))^2+\gamma^2\sin^2(ka)}}\nonumber\\
	&=& 0
\end{eqnarray}
bringing $x$ to one side, multiplying both sides by the denominator and squaring gives,
\begin{align}
	&x^2\hbar^2\left((g-\cos(ka))^2+\gamma^2\sin^2(ka)\right)\nonumber\\
	&=4a^2J^2t^2\left(g+(\gamma^2-1)\cos(ka)\right)^2\sin^2(ka) \ .
\end{align}
Replacing $\sin^2(ka)=1-\cos^2(ka)$, putting $z\equiv\cos(ka)$, and collecting as a quartic polynomial gives
\begin{align}\label{eq:FirstDeriv}
	0=&\;\left(\gamma ^2-1\right)^2 v_{I}^2 t^2 z^4+2 \left(\gamma ^2-1\right) g v_{I}^2 t^2 z^3\nonumber\\&+\left(g^2 v_{I}^2 t^2-\left(\gamma ^2-1\right)^2 v_{I}^2 t^2-\gamma ^2 x^2+x^2\right)z^2 \nonumber \\&
	+ \left(-2 \left(\gamma ^2-1\right) g v_{I}^2 t^2-2 g x^2\right)z+\gamma ^2 x^2-g^2 v_{I}^2 t^2 \nonumber \\ & +g^2 x^2
\end{align}
where the Ising velocity $v_{\text{\tiny{I}}}$ was defined in Eq.\ \eqref{eq:vIsing}.
The light cones correspond to the control parameter values where solutions coincide, that is, the stationary points of this equation. 

The vanishing of the second derivative of the generating function gives the equation
\begin{align}
& \frac{\partial^2\Phi}{\partial k^2}= \nonumber \\  & \frac{Jt\left(2a\gamma^2\cos(ka)\sin(ka)-2a(\cos(ka)-g)\sin(ka)\right)^2}{2\hbar\left((\cos(ka)-g)^2+\gamma^2\sin^2(ka)\right)^{3/2}}\nonumber\\
&-\frac{Jt\left(2a^2\gamma^2\cos^2(ka)-2a^2\cos(ka)(\cos(ka)-g)\right)}{\hbar\sqrt{(\cos(ka)-g)^2+\gamma^2\sin^2(ka)}} \nonumber \\
&-\frac{Jt\left(2a^2\sin^2(ka)-2a^2\gamma^2\sin^2(ka)\right)}{\hbar\sqrt{(\cos(ka)-g)^2+\gamma^2\sin^2(ka)}} =0 \ .
\end{align}
We now multiply both sides by $2\hbar\left((\cos(ka)-g)^2+\gamma^2\sin^2(ka)\right)^{3/2}/Jt$ and simplify,
\begin{align}
0=&\;\frac{1}{2}a^2\bigl(3(4g^2+(\gamma^2-1)^2)-2g(9+4g^2-5\gamma^2)\cos(ka)\nonumber\\
&+4(\gamma^4-1+g^2(2\gamma^2-3))\cos(2ka)\\
&+6g(\gamma^2-1)\cos(3ka)+(\gamma^2-1)^2\cos(4ka)\bigr) \ . \nonumber 
\end{align}
Next we make the replacements $\cos(2ka)=2\cos^2(ka)-1$; $\cos(3ka)=4\cos^3(ka)-3\cos(ka)$; and $\cos(4ka)=8\cos^4(ka)-8\cos^2(ka)+1$. Defining again $z\equiv \cos(ka)$, and dividing both sides by $a^2/2$,
\begin{align}\label{eq:SecondDeriv}
0=&\;8(\gamma^2-1)^2z^4+24g(\gamma^2-1)z^3-8(2\gamma^2(\gamma^2-1)\nonumber\\
&+g^2(2\gamma^2-3))z^2-8g(g^2+\gamma^2)z+8\gamma^2(g^2+\gamma^2-1) \ .
\end{align}
The light cones/caustics correspond to simultaneous solutions of Eqns.\ \eqref{eq:FirstDeriv} and \eqref{eq:SecondDeriv} and hence correspond to the Lieb-Robinson (LR) bound which is the solution which maximizes the propagation speed of the quasiparticles. 

In the next section we describe how the triple coalescence of stationary points give rise to the Pearcey function which provides the inner cone in Fig.\ \ref{fig:DoubleCone}. The three stationary points which coalesce are those between the dashed lines in Fig.\ \ref{fig:XYDispersion}. For $0<\gamma<1$ and $0<g<1$, this coalescence occurs at $k=0$, thus $z=1$, and Eq.\ \eqref{eq:SecondDeriv} yields solutions $g=1$ and $g=1-\gamma^2$. The $g=1$ solution is highly singular for nonzero anisotropy, while the solution $g=1-\gamma^2$ is the key for triple root coalescence.

\subsection{\label{Appdx:Pearcey}Diffraction integral for the cusp wave catastrophe}

Let us begin by defining the Pearcey function which is the canonical form of the wavefunction corresponding to the cusp catastrophe. The definition of the Pearcey function that we use is
\begin{equation}\label{eq:Pearcey}
\mathrm{Pe}(C_1,C_2)\equiv \frac{1}{2\pi}\int_{-\infty}^{\infty}\mathrm{d}s\;\mathrm{e}^{-\mathrm{i}\left(C_1s+\frac{C_2}{2}s^2+\frac{s^4}{4}\right)}\;.
\end{equation}
It features two parameters $C_{1}$ and $C_{2}$ and is generally a complex function. In fact, the common definition of the Pearcey function is the complex conjugate of \eqref{eq:Pearcey}, however for our purposes the above definition is more convenient.

Since the coalescence of extrema in $\Phi$ occurs at $k=0$, we expand to fourth-order, and factor out $J/\hbar$ which we will later use for scaling,
\begin{align}
\Phi(k;J)\approx&\;\frac{J}{\hbar}\biggl[2t(g-1)+\frac{2gax}{v_{\text{\tiny{I}}}}k\nonumber\\&+\frac{1}{2}\frac{2a^2t(\gamma^2+g-1)}{(g-1)}k^2
-\frac{1}{4}4a^4t\Gamma k^4\biggr]
\end{align}
where we have defined the following parameter,
\begin{equation}
\Gamma\equiv\frac{(g^3-1-2\gamma^2+3\gamma^4+g(3-2\gamma^2)+g^2(4\gamma^2-3))}{12(g-1)^3}\;.
\end{equation}
Note that the solution $g=1-\gamma^2$ will kill off the quadratic piece.

We now rescale our integration variable,
\begin{equation}
s=\sqrt{2}a\left(t\Gamma\right)^{\frac{1}{4}}k
\end{equation}
then our wavefunction locally takes the form,
\begin{align}\label{eq:PearceyLikeA}
\Psi_{\mathrm{Pe}}(C_1,C_2;J)\approx&\;\frac{\mathrm{e}^{\mathrm{i}\theta(t)}}{2\pi}\sqrt{\frac{J}{\hbar v_{\text{\tiny{I}}}}}\left(\frac{\gamma^2+g-1}{(g-1)C_2}\right)^{\frac{1}{2}}\nonumber\\
&\times\int_{-S}^{S}\mathrm{d}s\;\mathrm{e}^{-\mathrm{i}\frac{J}{\hbar}\left(C_1s+\frac{C_2}{2}s^2+\frac{s^4}{4}\right)}
\end{align}
with definitions,
\begin{align}
C_1(g,\gamma;x,t)=&\;-\frac{\sqrt{2}x}{v_{\text{\tiny{I}}}\left(t\Gamma\right)^{\frac{1}{4}}}\label{eq:ScaledC1}\\
C_2(g,\gamma;x,t)=&\;-\frac{\gamma^2+g-1}{g-1}\left(\frac{t}{\Gamma}\right)^{\frac{1}{2}}\label{eq:ScaledC2} 
\end{align}
and integration limit,
\begin{align}
S=&\;\sqrt{2}\pi\left(t\Gamma\right)^{\frac{1}{4}}  \ .
\end{align}
Eq.\ \eqref{eq:PearceyLikeA} shows that the wavefunction for the inner cone can locally be expressed as a diffraction integral which is generated by the cusp catastrophe $\Phi_2=C_1s+C_2s^2/2+s^4$, and is thus directly related to the canonical Pearcey function when $t$ is reasonably large and $J/\hbar = 1$ (below we will see that we can choose any value of  $J/\hbar$ and it will simply rescale the coordinates). Note, however, that the normalization of the wavefunction restricts the bounds of the integral as $t\to 0$, and so no true cusp point can occur at the origin since $\Phi$ also vanishes there. 
Nevertheless, the region of integration is proportional to $t^{1/4}$ and so is larger than the separation between the stationary points as $t\to 0$ since for any quartic equation of the form $\Phi_2$ the position of the stationary points in the $s$ coordinate is proportional to $\sqrt{C_2}$ so that for any infinitesimal time $\mathrm{d}t$ the separation between them is proportional to only $(\mathrm{d}t)^{1/2}$. Thus it becomes imperative that we consider the effects of all three stationary points, giving rise to the Pearcey-like function described in Eq.\ \eqref{eq:PearceyLike}.

Finally, in order to keep our expressions consistent for $|g|<1$ and $|g|>1$, we can instead factor out $Jg/\hbar$ overall. The above results are then identical up to a factor of $1/g$, which can be absorbed into $s$ and is irrelevant for the scaling. Thus, the expression $v_{\text{\tiny{I}}}$ given in Eq.\ \eqref{eq:vIsing} may be used in Eq.\ \eqref{eq:ScaledC1} generally.

\subsection{Self-similar scaling of the cusp wave catastrophe}

Now we scale the coupling strength, which corresponds to the width of the dispersion relation, from $J\to J'$. As we do so we enforce $Js^4=J's'^4$ so that the wavefunction maintains its basic form. Then, the Berry scaling is,
\begin{equation}\label{eq:C1Scaling}
JC_1s=JC_1\left(\frac{J'}{J}\right)^{\frac{1}{4}}s'=J'C_1\left(\frac{J}{J'}\right)^{\frac{3}{4}}s'\;,
\end{equation}
and,
\begin{equation}\label{eq:C2Scaling}
J\frac{C_2}{2}s^2=J\frac{C_2}{2}\left(\frac{J'}{J}\right)^{\frac{1}{2}}s'^2=J'\frac{C_2}{2}\left(\frac{J}{J'}\right)^{\frac{1}{2}}s'^2\;,
\end{equation}
with Arnol'd scaling given by,
\begin{equation}\label{eq:ArnoldScaling}
\sqrt{J}\mathrm{d}s=\sqrt{J}\left(\frac{J'}{J}\right)^{\frac{1}{4}}\mathrm{d}s'=\sqrt{J'}\left(\frac{J}{J'}\right)^{\frac{1}{4}}\mathrm{d}s'\;.
\end{equation}
These are the scaling factors for the cusp wave catastrophe as listed in Table \ref{tab:catastrophetable}. As we tune $J$, it is convenient to keep the caustic in the same place. This is done by simultaneously tuning $a$ such that the Ising velocity $v_{\text{\tiny{I}}}$ is constant.

\section{\label{Appdx:TFIMCaustics}Caustics in the Transverse-Field Ising Model}

As mentioned in the main text, the outer light cone in the XY model is dominated by its Airy-like behaviour because it arises from the coalescence of just two stationary points. Since this also occurs in the simpler TFIM  (which is obtained by setting $\gamma=1$) we focus on this case here.

\subsection{Calculation of classical caustics in the TFIM}

As shown above for the cusp catastrophe case, we must first calculate the two caustic conditions
\begin{equation}\label{eq:CausticCond1}
\frac{\partial\Phi}{\partial k}=0=x-\frac{2agJt\sin(ka)}{\hbar\sqrt{g^2-2g\cos(ka)+1}}
\end{equation}
and
\begin{align}\label{eq:CausticCond2}
\frac{\partial^2\Phi}{\partial k^2}=0=&-\frac{2a^2gJt\cos(ka)}{\hbar\sqrt{g^2-2g\cos(ka)+1}}\nonumber\\
&+\frac{2a^2g^2Jt\sin^2(ka)}{\hbar(g^2-2g\cos(ka)+1)^{3/2}}
\end{align}
which must be simultaneously fulfilled. Rearranging Eq.\ \eqref{eq:CausticCond2},
\begin{align}
g(1-\cos^2(ka))=&\;\cos(ka)(g^2-2g\cos(ka)+1)\label{eq:LCEqn}
\end{align}
leading to $\cos(ka)=g$ or $\cos(ka)=1/g$, as expected. Inputting this into Eq.\ \eqref{eq:CausticCond1}, along with $\sin(ka)=\sqrt{1-g^2}$ (or $\sin(ka)=\sqrt{1-1/g^2}$ for $g>1$), we can solve for the LR velocity, which is identical to the Ising velocity we defined in the previous section,
\begin{equation}
v_{\text{\tiny{LR}}}=v_{\text{\tiny{I}}}\;.
\end{equation}

\begin{figure*}[t]
	\hspace{1.2cm}\textbf{DISCRETE WAVEFUNCTION}\hspace{3.0cm}\textbf{CONTINUUM APPROXIMATION}\\\vspace{0.1cm}
	\includegraphics[height=31.75mm]{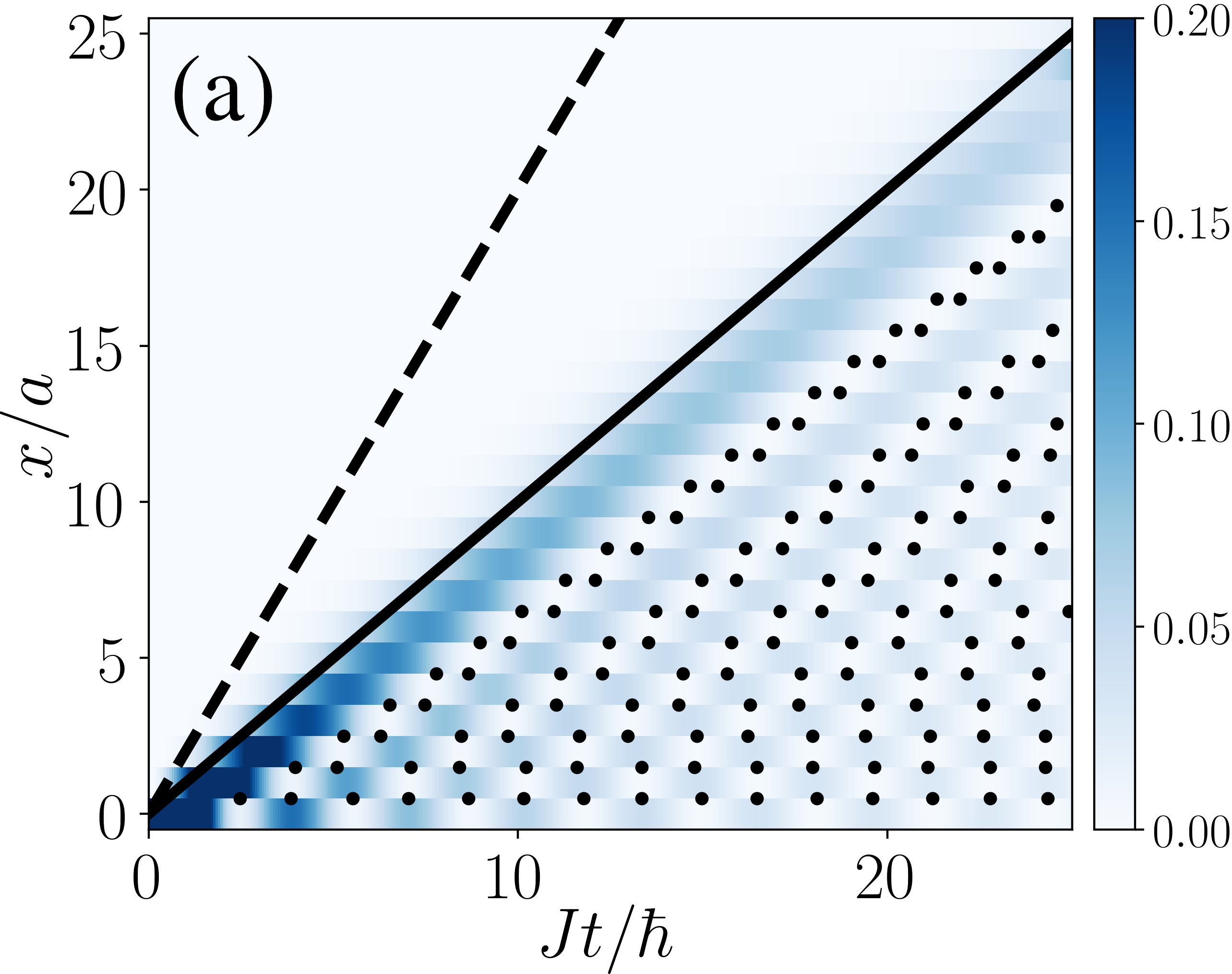}\hspace{0.5mm}
	\includegraphics[height=31.75mm]{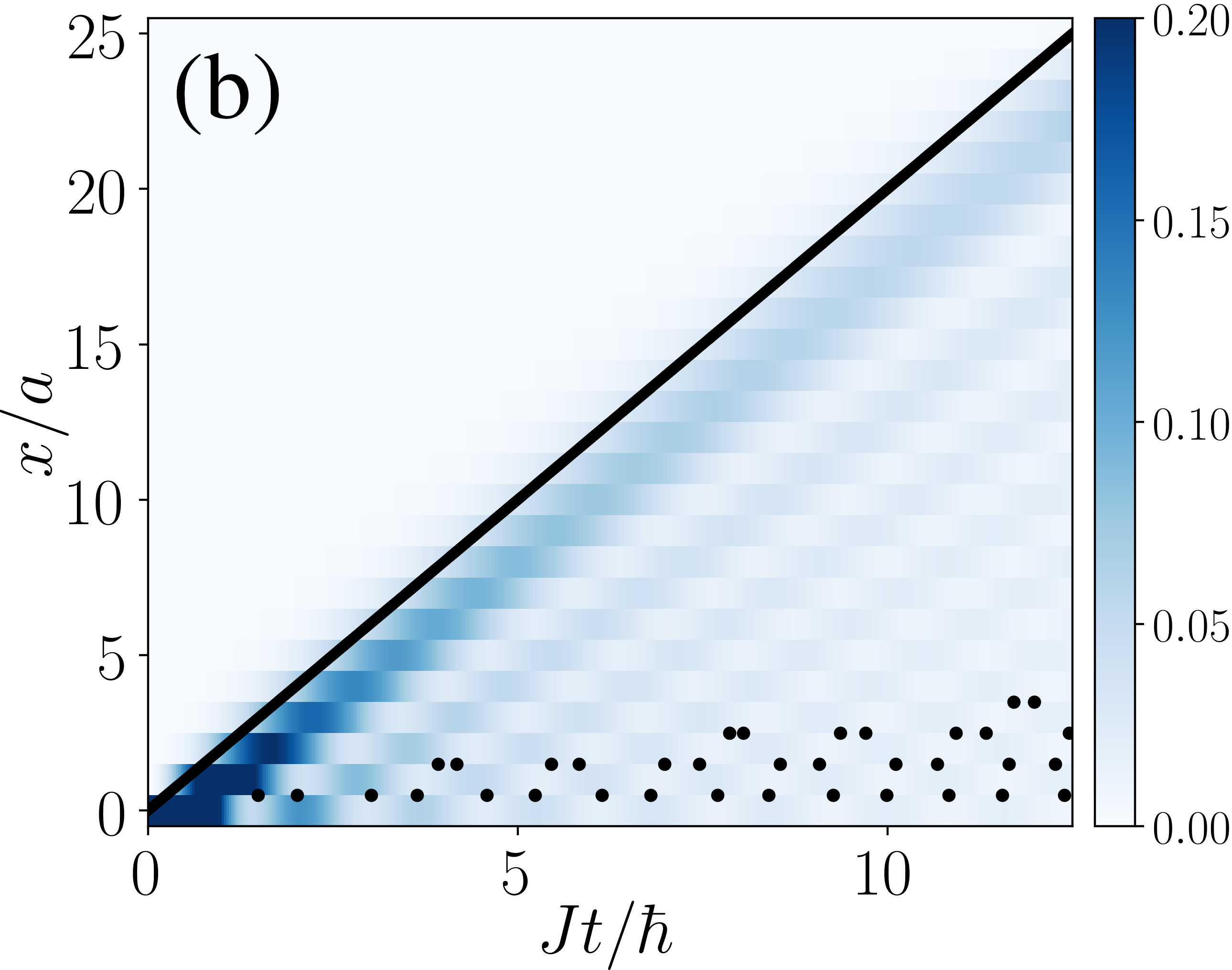}\hspace{0.8mm}
	\includegraphics[height=31.75mm]{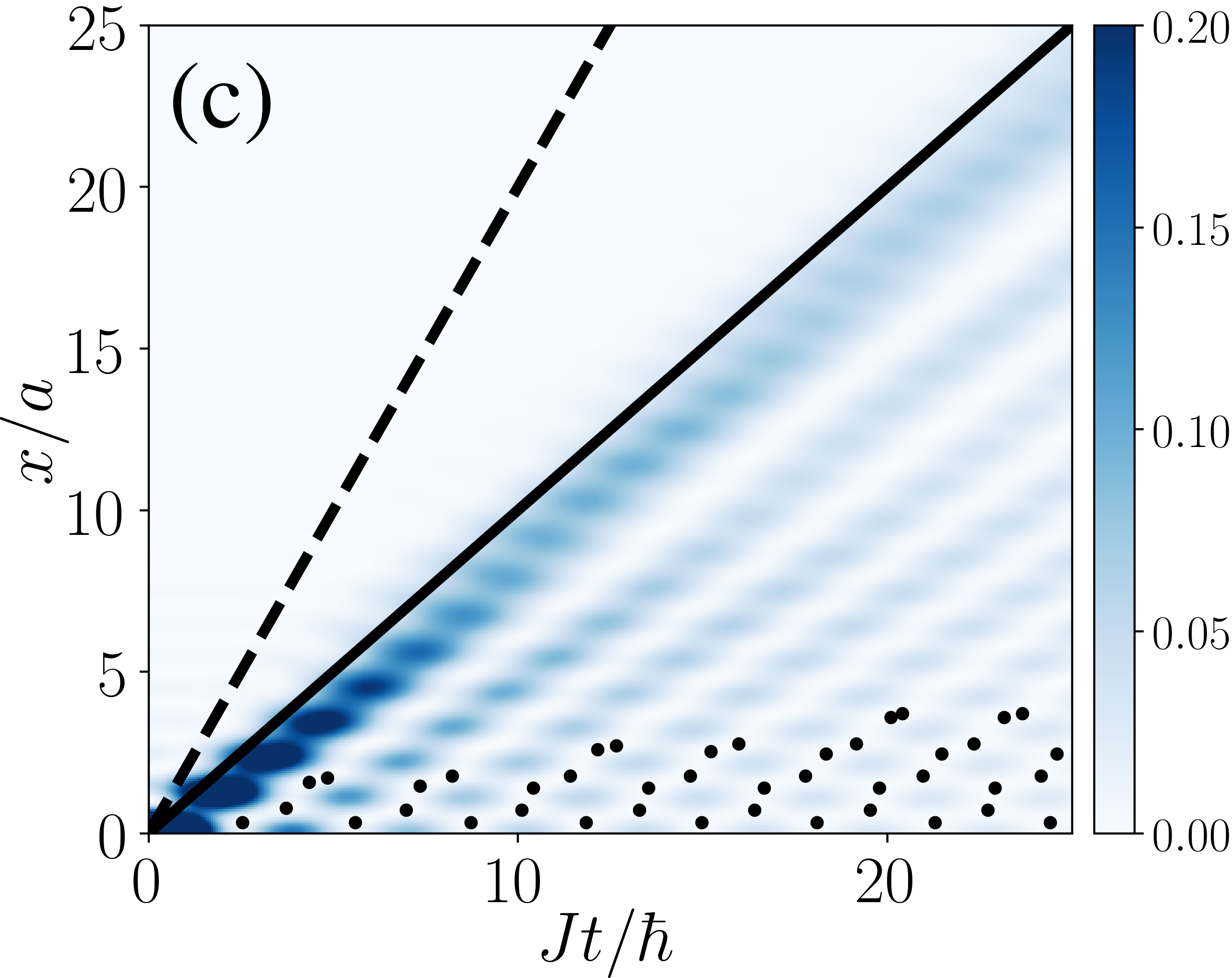}\hspace{0.5mm}
	\includegraphics[height=31.75mm]{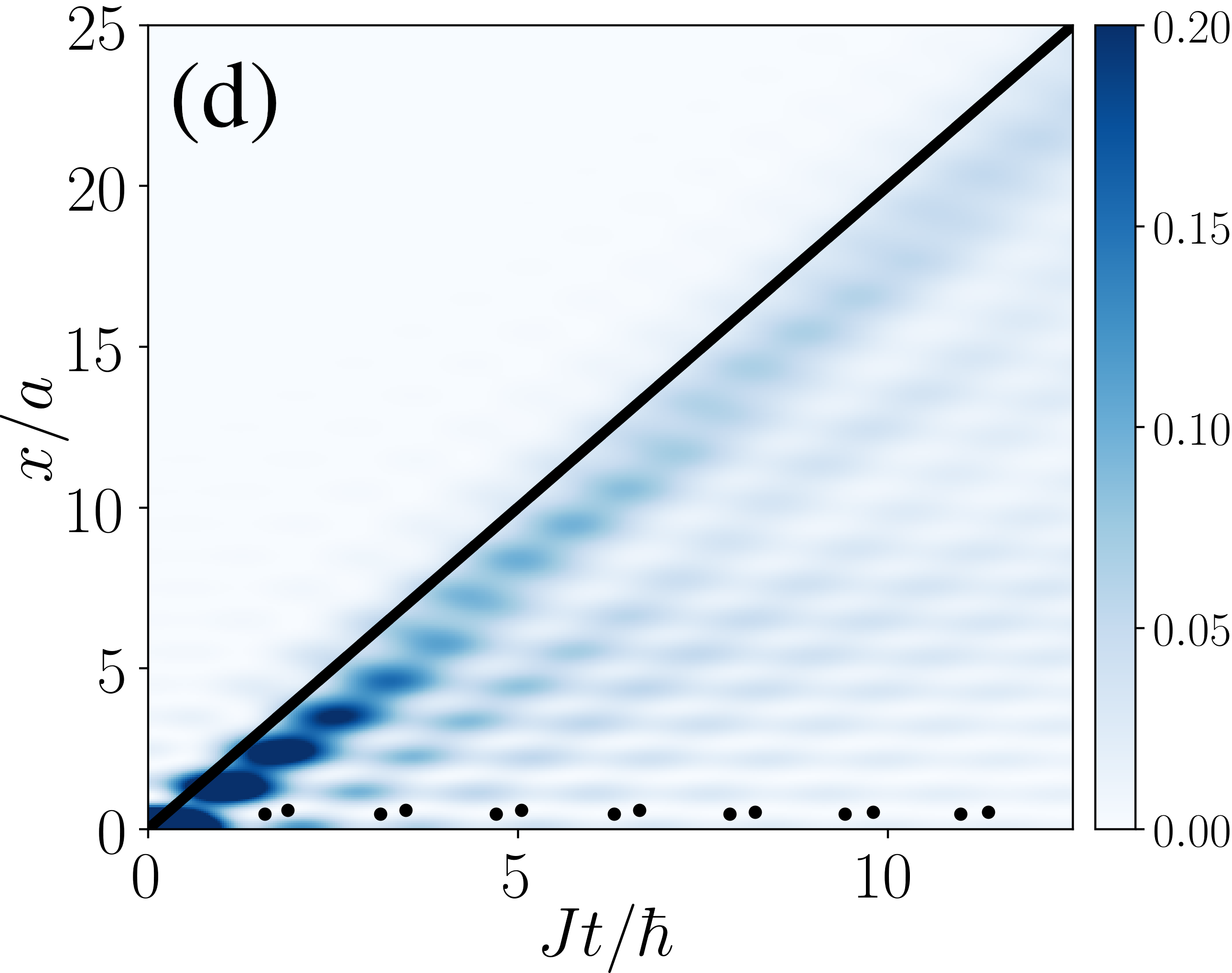}\\
	\includegraphics[height=31.75mm]{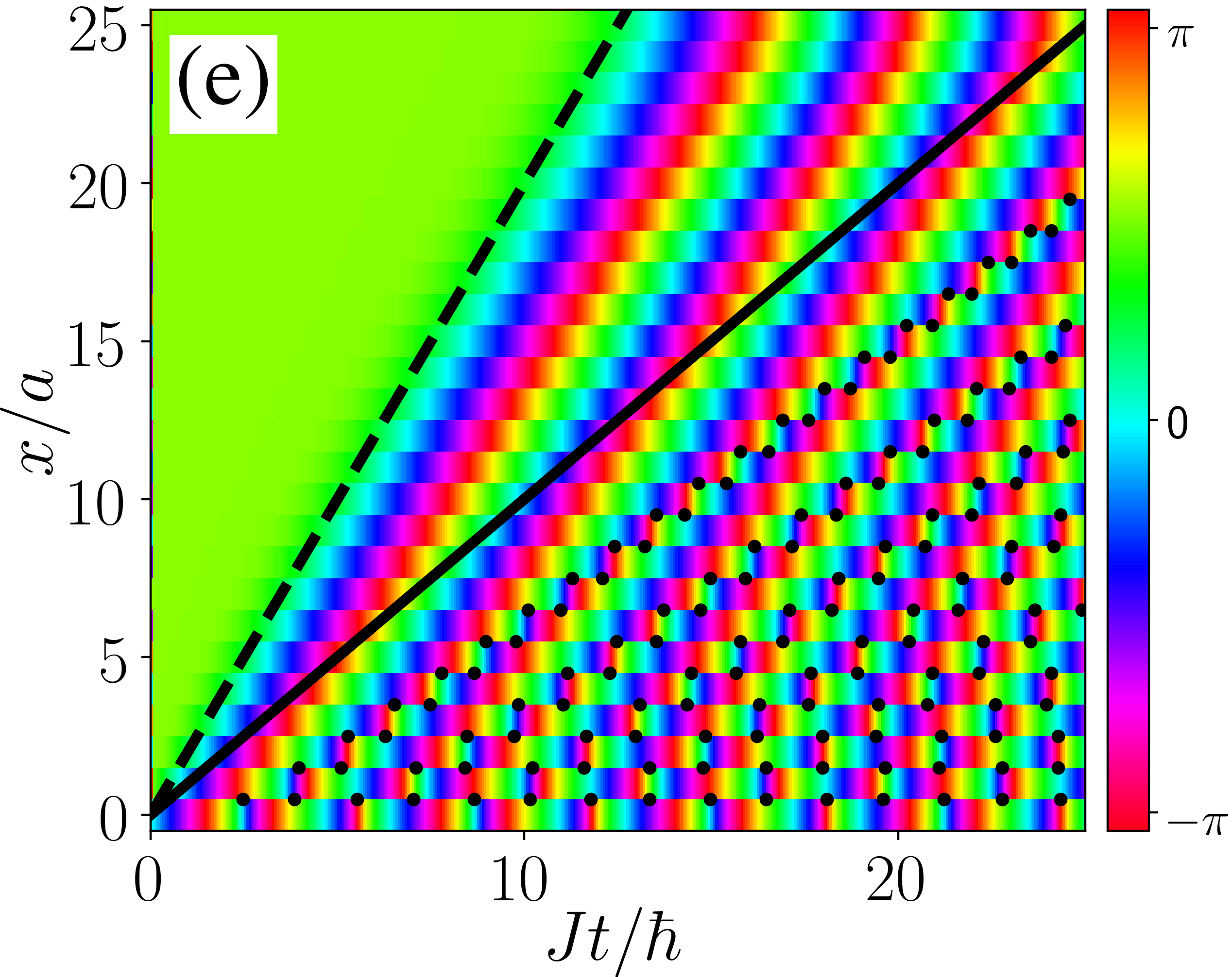}\hspace{0.5mm}
	\includegraphics[height=31.75mm]{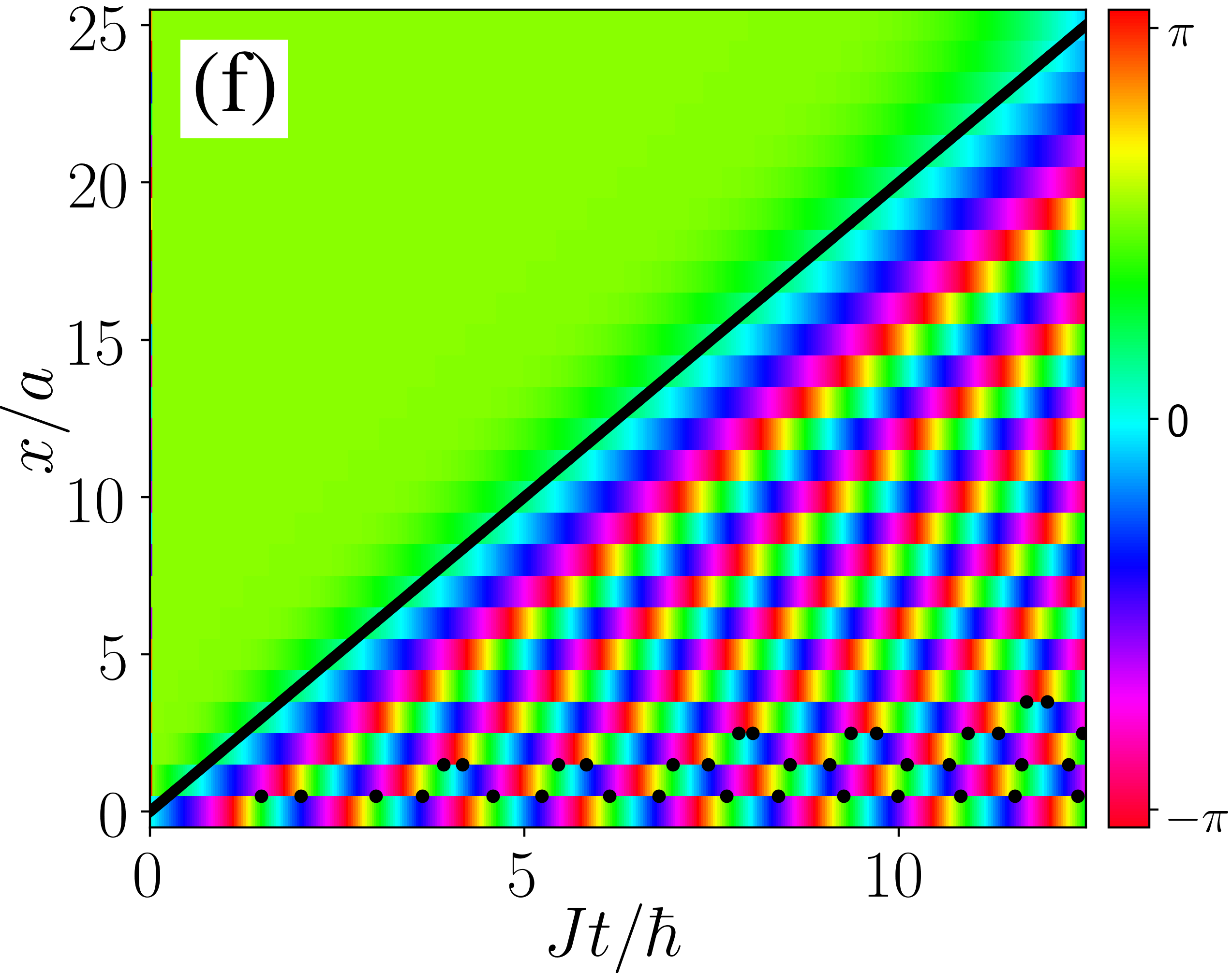}\hspace{0.8mm}
	\includegraphics[height=31.75mm]{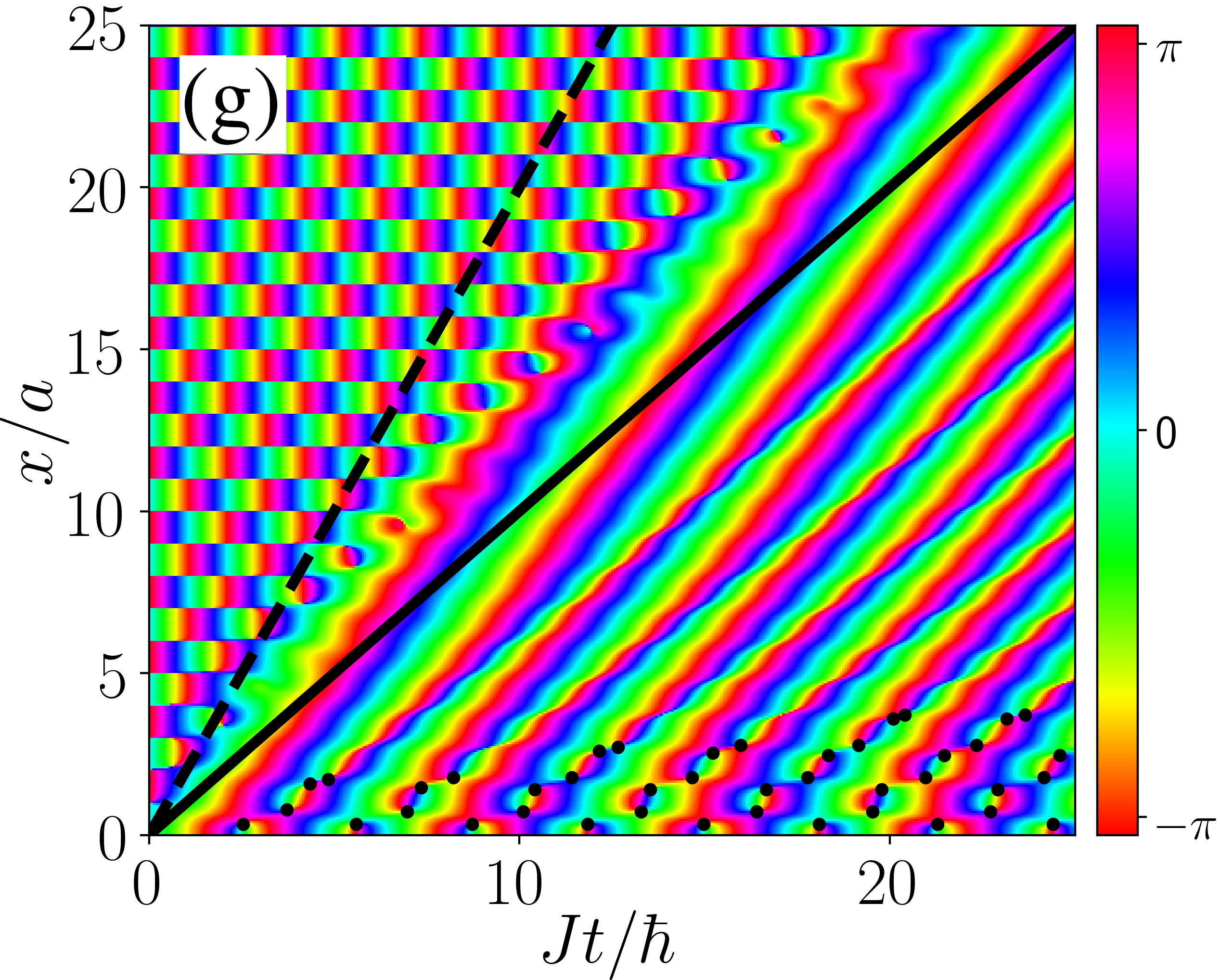}\hspace{0.5mm}
	\includegraphics[height=31.75mm]{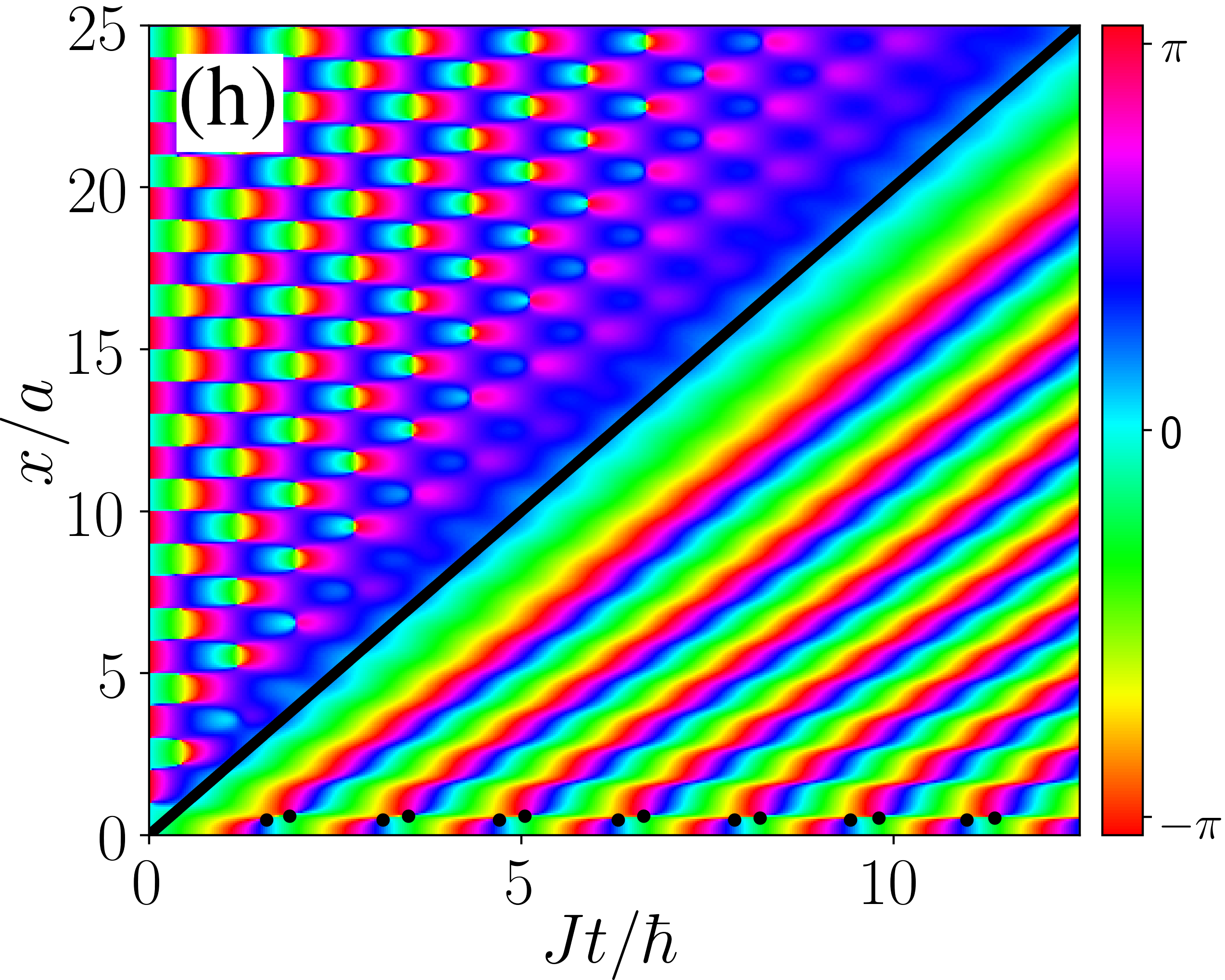}
	\caption{Caustics and vortices in the TFIM: discrete (exact) versus continuum approximation.  The initial condition is a single Bogoliubov fermion created at the centre of the chain (only half the chain is shown). The discrete wavefunction is given by Eq.\ \eqref{eq:PSISum} whereas the continuum approximation is given by Eq.\ \eqref{eq:Integral}.  \textbf{Top row [Panels (a)-(d)]:}  Amplitude of the wavefunction for $g=0.5\text{ and }1$. \textbf{Bottom row [Panels (e)-(h)]:}  phase of the same wavefunctions (corresponding to the panels directly above). The caustics are shown as solid black lines, while the imaginary caustics are plotted as dashed lines. The black dots mark the locations of vortices.}
	\label{fig:LightCones2}
\end{figure*}

Although the caustic lines are determined by the real solutions to Eq.\ \eqref{eq:LCEqn}, there exist imaginary solutions for which the Lieb-Robinson velocity $g$ designations are reversed. This seems to be responsible for lines of constant phase across the caustic (see Fig.\ \ref{fig:LightCones2}). The presence of two separate speeds within the light cone is also demonstrated by Cevolani \textit{et al.} in Ref.\ \cite{Cevolani2017}. We term these imaginary solutions as `imaginary caustics'.

\subsection{\label{Appdx:Airy}Diffraction  integral for the fold wave catastrophe}

The canonical wave catastrophe corresponding to the fold catastrophe is the Airy function. The definition of the Airy function that we use is
\begin{equation}\label{eq:Airy}
\mathrm{Ai}(C)\equiv \frac{1}{2\pi}\int_{-\infty}^{\infty}\mathrm{d}s\;\mathrm{e}^{\mathrm{i}(Cs+s^3/3)} \ .
\end{equation}
It features a single parameter $C$ and is a real function if $C$ is real.

 The stationary points of $\Phi$ coalesce when $k=(1/a)\arccos(g)$ for $g<1$ and $k=(1/a)\arccos(1/g)$ for $g>1$, respectively. Thus, for each of these cases, we will expand about these particular $k$ values to third order and factoring out $J/\hbar$ overall,
\begin{align}
\Phi(k;J)\approx&\; \frac{J}{\hbar}\left(-2t\sqrt{1-g^2}+\frac{x\hbar}{aJ}\arccos(g)\right)\nonumber\\
&+\frac{J}{\hbar}\left(\frac{x\hbar}{aJ}-2gt\right)\left(ka-\arccos(g)\right)\\
&+\frac{J}{\hbar}\frac{1}{3}gt\left(ka-\arccos(g)\right)^3\nonumber
\end{align}
for $g<1$ and,
\begin{align}
\Phi(k;J)\approx&\;  \frac{J}{\hbar}\left(-2t\sqrt{g^2-1}+\frac{x\hbar}{aJ}\arccos(1/g)\right)\nonumber\\
&+\frac{J}{\hbar}\left(\frac{x\hbar}{aJ}-2t\right)\left(ka-\arccos(1/g)\right)\\
&+\frac{J}{\hbar}\frac{1}{3}t\left(ka-\arccos(1/g)\right)^3\;,\nonumber
\end{align}
for $g>1$. Of course, the expansion will only capture the behaviour of the wavefunction close to the light-cone, however this is our primary objective. Furthermore, we are guaranteed that (up to a smooth change of variables) this cubic form in particular is structurally stable and will capture the qualitative features of $\Phi$.
We now rescale our integration variables as
\begin{align}
s_1^3=&\;gt\left(ka-\arccos(g)\right)^3\\
s_2^3=&\;t\left(ka-\arccos(1/g)\right)^3\;,
\end{align}
Thus,
\begin{align}
\Phi_{\mathrm{Ai}}(s;J)=&\;\frac{J}{\hbar}\biggl[\left(-2t\sqrt{1-g^2}+\frac{2gx}{v_{I}}\arccos(g)\right)\nonumber\\
&+2\left(\frac{x}{v_{I}}-t\right)\left(\frac{g^2}{t}\right)^{\frac{1}{3}}s_1+\frac{1}{3}s_1^3\biggr]\;,
\end{align}
for $g<1$ and,
\begin{align}
\Phi_{\mathrm{Ai}}(s;J)=&\;\frac{J}{\hbar}\biggl[\left(-2t\sqrt{g^2-1}+\frac{2x}{v_{I}}\arccos(1/g)\right)\nonumber\\
&+2\left(\frac{x}{v_{I}}-t\right)\left(\frac{1}{t}\right)^{\frac{1}{3}}s_2+\frac{1}{3}s_2^3\biggr]\;,
\end{align}
for $g>1$.

Now we define the control variable as
\begin{equation}
C=\begin{cases}
2\left(\frac{x}{v_{I}}-t\right)\left(\frac{g^2}{t}\right)^{\frac{1}{3}}&,\qquad g<1\\
2\left(\frac{x}{v_{I}}-t\right)\left(\frac{1}{t}\right)^{\frac{1}{3}}&,\qquad g>1\;,
\end{cases}
\end{equation}
so that,
\begin{equation}
\Psi(C;J)=\begin{cases}
\frac{e^{\Theta_1}}{2\pi(gt)^{\frac{1}{3}}\sqrt{a}}\int_{s^{\text{\tiny{Min}}}}^{s^{\text{\tiny{Max}}}}\mathrm{d}s\;e^{\frac{\mathrm{i}J}{\hbar}\Phi_1(C,s)}&,\qquad g<1\\
\frac{e^{\Theta_2}}{2\pi t^{\frac{1}{3}}\sqrt{a}}\int_{s^{\text{\tiny{Min}}}}^{s^{\text{\tiny{Max}}}}\mathrm{d}s\;e^{\frac{\mathrm{i}J}{\hbar}\Phi_1(C,s)}&,\qquad g>1\;,
\end{cases}
\end{equation}
with,
\begin{equation}
\Phi_1(C,s)=Cs+\frac{1}{3}s^3\;,
\end{equation}
\begin{equation}
\frac{1}{\sqrt{a}}ds=\begin{cases}
\sqrt{\frac{2gJ}{v_{I}\hbar}}ds&,\qquad g<1\\
\sqrt{\frac{2J}{v_{I}\hbar}}ds&,\qquad g>1\;,
\end{cases}
\end{equation}
and limits,
\begin{align}
s^{\text{\tiny{Min}}}=&\begin{cases}
\;\left(gt\right)^{\frac{1}{3}}\left(-\pi-\arccos(g)\right)&\qquad g<1\\
\;t^{\frac{1}{3}}\left(-\pi-\arccos(1/g)\right)&\qquad g>1
\end{cases}\\
s^{\text{\tiny{Max}}}=&\begin{cases}
\;\left(gt\right)^{\frac{1}{3}}\left(\pi-\arccos(g)\right)&\qquad g<1\\
\;t^{\frac{1}{3}}\left(\pi-\arccos(1/g)\right)&\qquad g>1
\end{cases}\;.
\end{align}
Note that $s^{\text{\tiny{Min}}}<0$ and $s^{\text{\tiny{Max}}}>0$. Thus, if we assume long enough times, then it is reasonable to take these integration limits to plus and minus infinity. We now have a description of the wavefunction local to the light cone using a fold catastrophe integral, which in the limit of $J/\hbar \to 1$ will become the Airy integral.

\subsection{Self-similar scaling of the fold wave catastrophe}

As for the cusp case, we can extract the scaling properties of the fold wave catastrophe by considering the change from $J\to J'$. Under this transformation we assume that $Js^3=J's'^3$. Then,
\begin{equation}
JCs=JC\left(\frac{J'}{J}\right)^{\frac{1}{3}}s'=J'C\left(\frac{J}{J'}\right)^{\frac{2}{3}}s'\;,
\end{equation}
and,
\begin{equation}
\sqrt{J}ds=\sqrt{J}\left(\frac{J'}{J}\right)^{\frac{1}{3}}ds'=\sqrt{J'}\left(\frac{J}{J'}\right)^{\frac{1}{6}}ds'
\end{equation}
Taking the integral limits to infinity (long times),
\begin{equation}
\Psi_{\mathrm{Ai}}(C;J)\propto\sqrt{J'}\left(\frac{J}{J'}\right)^{\frac{1}{6}}\int_{-\infty}^{\infty}\mathrm{d}s'\;e^{\mathrm{i}\frac{J'}{\hbar}\left(\left(\frac{J}{J'}\right)^{\frac{2}{3}}Cs'+\frac{1}{3}s'^3\right)}
\end{equation}
or, equivalently,
\begin{equation}
\Psi_{\mathrm{Ai}}(C;J)=\left(\frac{J}{J'}\right)^{\frac{1}{6}}\Psi_{\mathrm{Ai}}\left(\left[\frac{J}{J'}\right]^{\frac{2}{3}}C;J'\right)\;.
\end{equation}
We have therefore obtained the scaling factors for the fold wave catastrophe as listed in Table \ref{tab:catastrophetable}.

\section{\label{Appdx:SpinFlip}Spin Flip State $\Psi_\mathrm{X}(x,t)$}

In this paper we mainly consider an initial state consisting of a single fermionic quasiparticle localized on a particular site. However, in Section \ref{sec:expt} we instead consider the initial state where all the spins are polarized along the $x$ direction except for the central spin which is flipped such that the time evolved wavefunction is
\begin{equation}\label{eq:XStateA}
\Psi_\mathrm{X}(x,t)\equiv\bra{x}\mathrm{e}^{-\mathrm{i}Ht/\hbar}\ket{\uparrow^x...\uparrow^x\downarrow^x\uparrow^x...\uparrow^x}\;.
\end{equation}
Because experiments with ions can easily address individual spins, and spins and quasiparticles are not quite the same thing, it important to consider this kind of state.

Evaluating the time evolution of spin chains is generally far simpler in the Bogoliubov basis. However,  to introduce physical spins we begin with the JW basis which is related to the Bogoliubov basis by the Bogoliubov rotation:
\begin{equation}
\tilde{c}_k^\dagger=u_k\tilde{b}_{k}^\dagger-\mathrm{i}v_k\tilde{b}_{-k}\; .
\end{equation}
We identify the creation of a JW fermion at the centre of the lattice as a spin flip from $\uparrow^x$ to $\downarrow^x$ via the inverse JW transformation:
\begin{equation}
c_{j}^\dagger=\left(\prod\limits_{i>j}\sigma_{i}^x\right)\sigma_j^-\;.
\end{equation}
It is also important to note that the JW and Bogoliubov vacuums are related by
\begin{equation}
\ket{0}_c=\prod\limits_{k>0}\left(u_{k}-\mathrm{i}v_{k}\tilde{b}_{k}^\dagger\tilde{b}_{-k}^\dagger\right)\ket{0}_b \ .
\end{equation}
Starting with the centre spin $(x=0)$ down,
\begin{align}
\hspace{-0.5cm}\ket{\Psi_0}=&\;c_{x=0}^\dagger\ket{0}=\sum_{k_1}\tilde{c}_{k_1}^\dagger\ket{0}\nonumber\\
=&\sum_{k_1}(u_{k_1}\tilde{b}_{k_1}^\dagger-\mathrm{i}v_{k_1}\tilde{b}_{-k_1}) \nonumber \\ & \times \prod\limits_{k_2>0}\left(u_{k_2}-\mathrm{i}v_{k_2}\tilde{b}_{k_2}^\dagger\tilde{b}_{-k_2}^\dagger\right)\ket{0}_b\;,
\end{align}
and using the following relation,
\begin{align}
&\tilde{b}_{k_1}^\dagger\prod\limits_{k_2>0}\left(u_{k_2}-\mathrm{i}v_{k_2}\tilde{b}_{k_2}^\dagger\tilde{b}_{-k_2}^\dagger\right)\ket{0}_b\nonumber\\
&=u_{k_1}\tilde{b}_{k_1}^\dagger\prod\limits_{k_2>0,\;|k_2|\neq |k_1|}\left(u_{k_2}-\mathrm{i}v_{k_2}\tilde{b}_{k_2}^\dagger\tilde{b}_{-k_2}^\dagger\right)\ket{0}_b,
\end{align}
we get,
\begin{align}
\ket{\Psi_0}=\sum_{k_1}\tilde{b}_{k_1}^\dagger\prod\limits_{k_2>0,\;|k_2|\neq |k_1|}\left(u_{k_2}-\mathrm{i}v_{k_2}\tilde{b}_{k_2}^\dagger\tilde{b}_{-k_2}^\dagger\right)\ket{0}_b\;.
\end{align}
Next, we evolve in time using a more convenient representation of the time-evolution operator,
\begin{equation}
\mathrm{e}^{-\frac{\mathrm{i}t}{\hbar}\sum\limits_{k}\epsilon_{k}\left(\tilde{b}_{k}^\dagger\tilde{b}_{k}-\frac{1}{2}\right)}=\mathrm{e}^{\mathrm{i}\theta(t)}\prod_{k}\left(1-\left(1-\mathrm{e}^{-\frac{\mathrm{i}t\epsilon_{k}}{\hbar}}\right)\tilde{b}_{k}^\dagger\tilde{b}_{k}\right) \ .
\end{equation}
Dropping the global phase factor, and projecting this state onto real space $\Psi(x_i,t)=\braket{x_i|\Psi(t)}$ using
\begin{align}
\bra{x_i}= &\bra{0}_cc_i  \nonumber \\ =&\;\sum_{k_3}\mathrm{e}^{\mathrm{i}k_3x_i}\bra{0}_b\prod\limits_{\substack{k_4>0\\|k_4|\neq |k_3|}}\left(u_{k_4}+\mathrm{i}v_{k_4}\tilde{b}_{-k_4}\tilde{b}_{k_4}\right)\tilde{b}_{k_3} \ ,
\end{align}
we arrive at, after a fair amount of algebra,
\begin{align}
\Psi(x_i,t) & =\sum_{k_1}\mathrm{e}^{-\frac{\mathrm{i}t\epsilon_{k_1}}{\hbar}}\mathrm{e}^{\mathrm{i}k_1x_i}\prod\limits_{\substack{k_2>0\\|k_2|\neq|k_1|}}\left(u_{k_2}^2+v_{k_2}^2\mathrm{e}^{-\frac{\mathrm{i}t\epsilon_{k_2}}{\hbar}}\right)\nonumber\\
+ & \sum_{k_1}\mathrm{e}^{-\frac{2\mathrm{i}t\epsilon_{k_1}}{\hbar}}\mathrm{e}^{\mathrm{i}k_1x_i}v_{k_1}^2\prod\limits_{\substack{k_2>0\\|k_2|\neq|k_1|}}\left(u_{k_2}^2+v_{k_2}^2\mathrm{e}^{-\frac{\mathrm{i}t\epsilon_{k_2}}{\hbar}}\right)\;.
\end{align}

\begin{figure}[t]
	\centering
	\includegraphics[height=50mm]{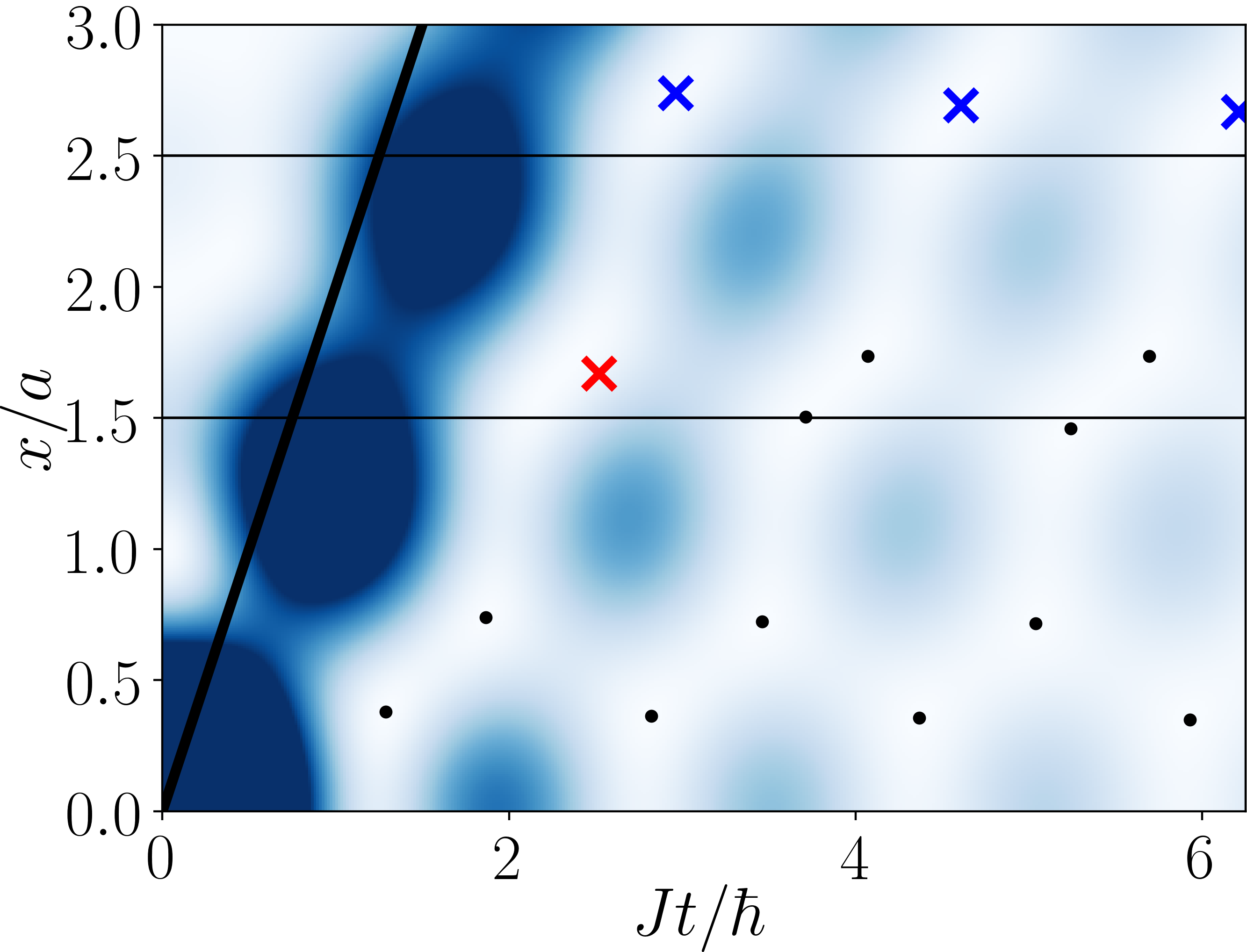}
	\caption{Graphic depicting vortex annihilation in the CA for the TFIM. Here, $g=1.75$, and for clarity the lines $x/a=1.5$ and $x/a=2.5$ have been drawn. As $g$ is tuned toward the transition, vortex-antivortex pairs (black dots) will approach one another and eventually annihilate at a particular point in space-time, denoted with an `X'. It is the vortices which annihilate close to $x/a=1.5$ that we refer to as `primary' (red) and those which annihilate close to $x/a=2.5$ we refer to as `secondary' (blue, all annihilated in this image). In principle, there exist rows of vortices beyond these, but here we focus on those closer to the centre of the lattice and short times.}
	\label{fig:Annihilations}
\end{figure}

\section{Vortex Scaling}
\label{Appdx:vortexscaling}

Returning to our original initial condition of a single Bogoliubov fermion created at $x=0$, we can identify space-time vortices in the time evolved system. The discrete (exact) and CA results are compared for the TFIM in Fig.\ \ref{fig:LightCones2} where the same general trend is observed in both cases: fewer vortices at the QCP at $g=1$ than away from it at $g=0.5$. The vortices that survive at the QCP are those near to the centre of the chain at $x=0$, i.e.\ those closest to the position of the original excitation. In fact, in the CA only a single line of vortices on each side of the centre line survives. 

The vortices are located by breaking the light cone up into small loops and integrating the phase of the wavefunction around each one. For a loop containing a single vortex,
\begin{equation}
\int_\mathcal{C}\mathrm{d}\chi = \pm 2\pi 
\end{equation}
where the plus sign signifies a vortex and the minus sign an antivortex. For the discrete wavefunction the integral along the spatial part of the path $\mathcal{C}$ is replaced by a sum.

\begin{figure}[h!]
	\centering
	\includegraphics[height=46mm]{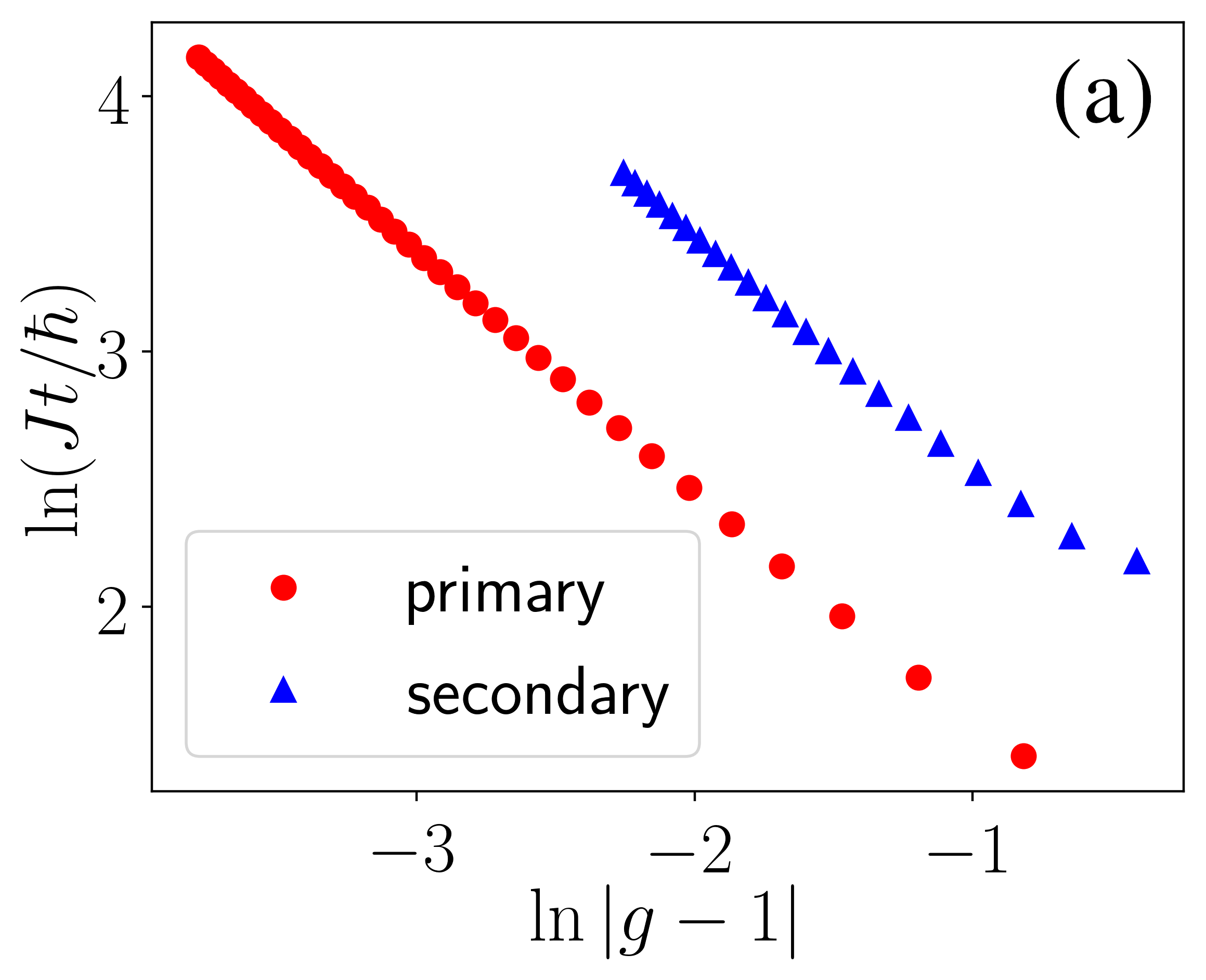}
	\includegraphics[height=46mm]{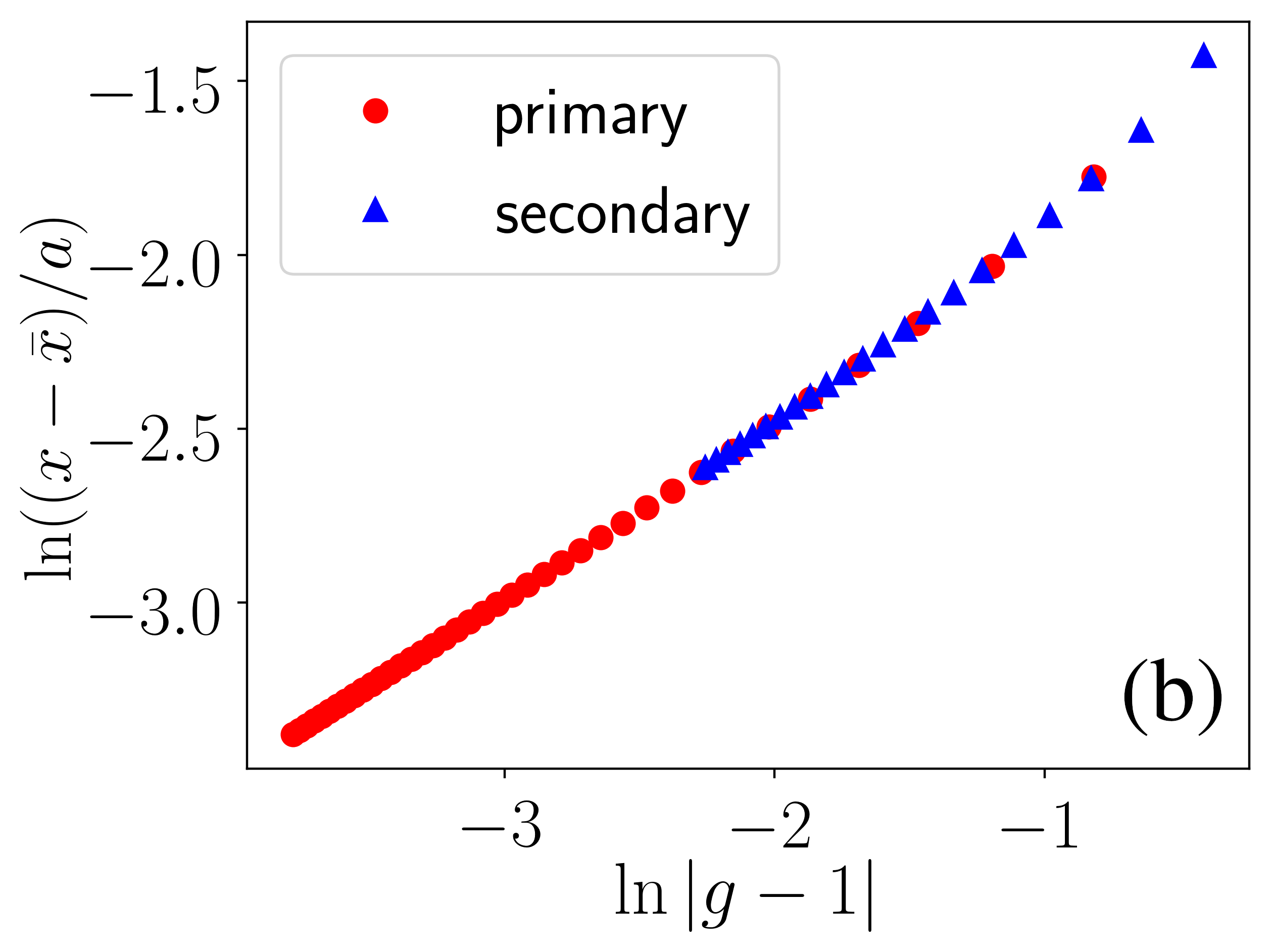}
	\caption{Vortex annihilation scaling in the TFIM within the CA. \textbf{Panel (a): } The time at which vortex annihilation occurs along a particular set of vortices will diverge as we approach the QCP. \textbf{Panel (b): } Each consecutive vortex pair will annihilate at a point in space ($x$) which approaches the midpoint between two lattice sites. Thus $\bar{x}=1.5a$ for the set of primary vortices, and $\bar{x}=2.5a$ for the secondary vortices. }
	\label{fig:Scaling}
\end{figure}

If we track the positions of the vortices as $g$ is varied we find that they flow in space-time in such a way that as the QCP is approached vortices and antivortices annihilate in pairs, each pair annihilating at a different point ($x,t$). This process is easier to follow in the CA than the discrete case because the discreteness in the lattice direction obscures the spatial location of vortices, so in this Appendix we specialize to the CA case (whereas the data presented in Fig.\ \ref{fig:Fig6} in the main text are for the discrete case). In particular, Fig.\  \ref{fig:Annihilations} gives a pictorial representation of the annihilations occurring near the centre of the lattice for $g=1.75$. We see that vortex-antivortex pairs converge on horizontal lines (i.e.\ spatial points) located at $x/a=\pm0.5,\pm1.5,\pm2.5,\ldots$.

The temporal behavior of the vortices can also be seen in Fig.\ \ref{fig:Annihilations}. For values of $g$ close the QCP the vortex-antivortex pairs that occur at short times annihilate and so never occur, or, said another way, as $g \to 1$ the creation time for vortex-antivortex pairs diverges, an example of critical slowing. Thus, there are two dimensions along which one can observe critical scaling: along $t$ and along $x$, and the data for these two directions are shown in Fig.\  \ref{fig:Scaling}. It is clear from the way that the data falls onto straight lines on a log-log scale as $g \to 1$ that the vortices display critical scaling. The figure shows two `sets' of vortices, where each set annihilates within a small region of $x/a$ at diverging time scales. The vortices we call primary vortices annihilate at positions approaching $\bar{x}=1.5a$, while $\bar{x}=2.5a$ for the secondary vortices. In the main text, we focus only on the primary vortices, since a greater number annihilate earlier in time and thus result in a less oscillatory integrand, allowing us to get closer to the transition while maintaining accuracy for a larger number of data points, but we see that the secondary vortices obey the same scaling.  The temporal scaling shown in Fig.\ \ref{fig:Scaling}(a) leads to a gradient of $-1$ and hence the relation $\nu z =1$ as explained in the main text [see inset in Fig.\ \ref{fig:Fig6}(b)]. The spatial scaling is shown in Fig.\ \ref{fig:Scaling}(b) and leads to a gradient of $0.5$.

\end{document}